\documentclass[usenatbib]{mn2e}
\usepackage{amsmath,amssymb,graphicx,aas_macros}

\begin{document}
\title{What determines satellite galaxy disruption?}
\author[Wetzel \& White]{Andrew R. Wetzel${}^{1}$, Martin White${}^{1,2}$\\
$^{1}$Department of Astronomy, University of California, Berkeley, CA 94720, 
USA\\
$^{2}$Department of Physics, University of California, Berkeley, CA 94720, USA
}

\date{July 2009}

\pagerange{\pageref{firstpage}--\pageref{lastpage}} \pubyear{2009}

\maketitle

\label{firstpage}

\begin{abstract}
In hierarchical structure formation, dark matter halos that merge with larger halos can persist as subhalos.
These subhalos are likely hosts of visible galaxies.
While the dense halo environment rapidly strips subhalos of their dark mass, 
the compact luminous material can remain intact for some time, making the correspondence of galaxies with severely stripped subhalos unclear.
Many galaxy evolution models assume that satellite galaxies eventually merge
with their central galaxy, but this ignores the possibility of satellite tidal disruption.
We use a high-resolution $N$-body simulation of cosmological volume to explore
satellite galaxy merging/disruption criteria based on dark matter subhalo
dynamics.
We explore the impact that satellite merging/disruption has on the Halo Occupation Distribution and radial profile of the remnants.
Using abundance matching to assign stellar mass/luminosity to subhalos, we compare with observed galaxy clustering, satellite fractions, cluster satellite luminosity functions, finding that subhalos reproduce well these observables.
Our results imply that satellite subhalos corresponding to $>0.2\,L_*$ galaxies must be well-resolved down to $1-3\%$ of their mass at infall to robustly trace the galaxy population.
We also explore a simple analytic model based on dynamical friction for satellite galaxy infall, finding good agreement with our subhalo catalog and observations.
\end{abstract}


\section{Introduction}

In the standard picture of galaxy formation, galaxies form at the centers of dark matter halos as baryons cool and contract toward the minimum of a halo's potential well \citep{WhiRee78,BluFabFlo86,Dub94,MoMaoWhi98}.
But halos are not isolated objects: they merge over time, and smaller halos can survive as substructure halos (subhalos) of larger halos after infall \citep{GhiMooGov98,TorDiaSye98,MooGhiGov99,KlyGotKra99}.
Thus, while galaxies form within distinct host halos at high redshift, as the Universe evolves galaxies then correspond directly to dark matter subhalos (and since galaxies can be centrals or satellites, in our terminology a ``subhalo'' refers to a satellite substructure or the central halo).

The evolution of subhalos involves more complicated dynamics than the formation of halos themselves.
Satellite subhalos experience severe mass stripping as they orbit, and dynamical friction causes their obits to sink toward halo center \citep{OstTre75}.
The evolution of galaxies is even more complex, involving gas dynamics, star formation, and various forms of feedback.
Thus, the precise evolutionary relation between galaxies and subhalos remains uncertain.
In particular, what is the relation between galaxy stellar mass and subhalo dark mass?
Do satellite galaxies experience appreciable star formation after infall?
How long after infall does a satellite galaxy survive, and what defines its final fate?
How much of its stellar mass is funneled into the central galaxy in a merger, as opposed to tidally disrupting into Intra-Cluster Light (ICL)?

Various prescriptions exist to map galaxies onto dark matter subhalos.
The simplest is based on subhalo abundance matching (SHAM), which assumes a monotonic relation between subhalo mass or circular velocity and galaxy stellar mass, populating subhalos such that one reproduces the observed stellar mass function \citep[SMF;][]{ValOst06,ConWecKra06,ShaLapSal06}.
Alternately, semi-analytic models track the star formation histories of subhalos across time with empirically motivated analytic prescriptions \citep[see][for a recent review]{Bau06}.
These methods all involve assumptions regarding satellite galaxy star formation after infall, the correspondence of galaxies with subhalos in the case of severe mass stripping, and the eventual fates of satellite galaxies.

Most models assume that all satellite galaxies eventually merge with their central galaxy.
Indeed, infalling satellites have long been thought to affect the mass and morphological evolution of central galaxies \citep{HauOst78}.
However, galaxy clusters are observed to contain an appreciable amount of stellar material in the form of ICL, though given its low surface brightness, constraints on the ICL vary considerably, from $5-50\%$ of the total cluster light \citep{LinMoh04,ZibWhiSch05,GonZarZab07,KriBer07}.
Additionally, ICL has been observed to contain significant structure, including tidal streams \citep{MihHarFel05}.
Thus, some satellite galaxies are at least partially, or perhaps entirely, disrupted into the ICL.

In simulations, the criteria for subhalo merging/disruption are influenced by mass and force resolution.
Insufficiently resolved subhalos will disrupt artificially quickly, leading to the problem of ``over-merging'' \citep{KlyGotKra99}.
A number of authors have examined in detail mass stripping from satellite subhalos over time \citep[e.g.,][]{MacColSpr09}, however, using a fixed mass resolution limit for all subhalos, regardless of their infall mass, leads to lower mass subhalos becoming (artificially) disrupted more quickly.
Cosmological $N$-body simulations are now attaining sufficient mass resolution to track subhalos through many orbits and extreme mass stripping, in some cases to masses much smaller than the luminous mass of the galaxies they would host.
At infall, the dark mass of a subhalo is an order of magnitude higher than its stellar mass, so tidal stripping of mass from subhalos cannot correspond directly to stellar mass stripping of its galaxy.
Thus, in using dark matter subhalos to track galaxies, one must be careful both to treat subhalo merging/disruption in a self-consistent manner not dependent merely on numerical resolution, and to calibrate the point at which the galaxies they host become merged/disrupted.

A variety of schemes have been used to define satellite galaxy merging/disruption via subhalos in simulations.
Under the assumption that a galaxies survives as long as its subhalo does,  \citet{WetCohWhi09a,WetCohWhi09b} use an absolute mass threshold for all subhalos, corresponding to their resolution limit, similar to \citet{KraBerWec04}, whose model also implies that subhalos coalesce if their centers come within $50\,h^{-1}$~kpc.
Several analyses assume that a satellite galaxy survives the disruption of its subhalo and eventually merges with its central galaxy by imposing an analytic infall timescale to a galaxy after subhalo disruption \citep{SprWhiTor01,KitWhi08,SarDeLDol08,MosSomMau09}.
Alternately, \citet{SteBulBar09} allow a satellite galaxy to merge before its subhalo is disrupted.
They assume that a galaxy's baryonic mass is coupled to the $10\%$ most bound mass of its subhalo and define satellite merging when a satellite subhalo has lost more than $90\%$ of its mass at infall.
\citet{YanMovdB09} consider a model where the survival probability of a satellite subhalo is a (decreasing) function of its infall mass to host halo mass ratio.

Other criteria have been used in purely semi-analytical models for the evolution of subhalos.
\citet{ZenBerBul05} and \citet{TayBab04} consider a satellite subhalo to be tidally disrupted when the mass of the subhalo has been stripped to a value less than the mass within $a_{\rm dis} r_s$, where $r_s$ is the subhalo's NFW \citep{NFW96} scale radius at infall, with $a_{\rm dis}=1$ \citep{ZenBerBul05} or $a_{\rm dis}=0.1$ \citep{TayBab04}.
Assuming a typical halo concentration of $c=10$, this implies that satellite subhalos will disrupt at $13\%$ or $0.3\%$ of their infall mass in the models, respectively -- an order of magnitude difference.
As an alternative to tidally disrupting, \citet{ZenBerBul05} consider the satellite subhalo merged with the center of its halo if it comes within $5$~kpc.
Applying the model of \citet{ZenBerBul05} to the build-up of the ICL, \citep{PurBulZen07} allow satellite disruption to begin at $\sim20\%$ of infall mass \citep[see also][for models of tidal disruption into the ICL]{MonMurBor06,ConWecKra07}.

Finally, mass loss criteria have also been applied to modelling dwarf-spheroidal galaxies in $N$-body simulations.
\citet{PenNavMcC08} find that dwarf satellites start to lose stellar mass to tidal stripping after $90\%$ dark mass loss, but full disruption occurs only after $>99\%$ dark mass loss.
\citet{MacKanFon09} find good agreement between their semi-analytic model and the Milky Way satellite luminosity function using $a_{\rm dis}=0.5-1$, corresponding to $5-13\%$ of the infall mass remaining.

Given the wide range in criteria for satellite merging/disruption, it is informative to explore the implications of various criteria and to test empirically which match well to the observed galaxy population.
Thus, in this paper we examine the extent to which galaxies can be mapped onto simulated dark matter subhalos throughout their evolution.
We assign galaxy stellar masses and luminosities to subhalos by matching their number densities to observed galaxy samples.
We examine the effects that different criteria for merging/disruption have on the satellite populations, including the HOD and radial distribution profile.
By comparing spatial clustering, satellite fractions, and cluster satellite luminosity functions to observations, we test the correspondence of galaxies to subhalos, and we constrain the criteria for satellite galaxy merging/disruption.

Before continuing, we clarify the possible fates of satellite galaxies.
The methods by which galaxies become removed from the satellite population are to (1) coalesce with the central galaxy, (2) coalesce with another satellite galaxy, (3) orbit outside their host halo, or (4) tidally disrupt into the ICL.
As noted above, many galaxy evolution models assume only (1).
As we describe below, our method for tracking subhalos in simulations intrinsically incorporates (2) and (3).
Thus, this paper focuses on methods for removing satellite galaxies via (1) and (4).
Since our tracking of dark matter subhalos does not unambiguously differentiate between these (and since both can occur simultaneously), we will use the term ``removal'' of satellite galaxies to indicate any mechanism that causes (1) or (4), which we refer to as ``merging'' and ``disruption'', respectively.

\section{Numerical Methods} \label{sec:numerical}
\subsection{Simulations and Subhalo Tracking} \label{sec:simulations}

To find and track halos and their subhalos, we use dark matter-only $N$-body simulations using the TreePM code of \citet{TreePM}.
To track subhalos with high resolution, we use a simulation of $1500^3$ particles in a periodic cube with side lengths $200\,h^{-1}$~Mpc.
For our $\Lambda$CDM cosmology ($\Omega_m=0.25$, $\Omega_\Lambda=0.75$, $h=0.72$, $n=0.97$ and $\sigma_8=0.8$), in agreement with a wide array of observations \citep{COBE,TegEisStr06,ACBAR,DunKomNol08}, this results in particle masses of $1.64 \times 10^8\,h^{-1}M_\odot$ and a Plummer equivalent smoothing of $3\,h^{-1}$~kpc.
Initial conditions were generated by displacing particles from a regular grid using 2LPT at $z=250$ where the RMS is $20\%$ of the mean inter-particle spacing.
We stored $45$ outputs evenly in $\ln(a)$ from $z=10$ to $0$, with an output time spacing of $\sim600\,$Myr at $z=0.1$, the main redshift of interest in this paper.
Given the limited volume of this simulation, and its implications for spatial clustering, we also use a separate simulation of $1500^3$ particles of size $720\,h^{-1}$~Mpc and a particle mass of $7.67\times 10^9\,h^{-1}M_\odot$ with the same cosmology and halo finder, though independent initial conditions.
We use this larger simulation to populate our subhalo catalog onto a more accurate halo mass function with more accurate large-scale clustering.
To examine the influence of cosmological parameters, we also populate our subhalo catalog into a simulation of $1024^3$ particles of size $500\,h^{-1}$~Mpc with $n=1.0$ and $\sigma_8=0.9$.
We explore the influence of simulation size and cosmology in the Appendix.

Our subhalo finding and tracking details are discussed extensively in \citet{WetCohWhi09a}\footnote{
While of similar mass resolution, the simulation here has larger volume and better force resolution than in \citet{WetCohWhi09a}.}, though we have implemented a number of improvements for better performance, as described below.
We find subhalos by first generating a catalog of halos using the Friends-of-Friends (FoF) algorithm \citep{DEFW} with a linking length of $b=0.168$ times the mean inter-particle spacing.\footnote{
The longer linking length of $b=0.2$ is often used, but it is more susceptible to joining together distinct, unbound structures and assigning a halo that transiently passes by another as a subhalo.}
We keep all groups that have more than $50$ particles, and halo masses quoted below are these FoF masses.
Within these ``(host) halos'' we then identify ``subhalos'' as gravitationally self-bound aggregations of at least $50$ particles bounded by a density saddle point, using a new implementation of the {\sl Subfind\/} algorithm \citep{SprWhiTor01}, where densities are smoothed over $32$ nearest neighbors.
The ``central'' subhalo is defined as the most massive subhalo within its host halo, and it includes all halo matter not assigned to ``satellite'' subhalos, which is typically $\sim90\%$ of the mass of the host halo.
Thus, every sufficiently bound halo hosts one central subhalo and can host multiple satellite subhalos.
A subhalo position is that of its most bound particle, and we define the halo center as that of its central subhalo. 

Each subhalo is given a unique child at a later time, based on its $20$ most bound particles.
We track subhalo histories across four consecutive outputs at a time since subhalos can briefly disappear during close passage with or through another subhalo.
In an improvement to the tracking method of \citet{WetCohWhi09a}, we now interpolate positions, velocities, and bound masses of these temporarily disappearing subhalos so that our tracking scheme does not artificially underestimate the satellite population.

Also, as discussed in \citet{WetCohWhi09a}, subhalo histories often include
cases of ``switches'': if a satellite becomes more massive than the central,
it becomes the central while the central becomes a satellite.
In these cases, there is typically not a well-defined single peak that
represents the center of the halo profile, and the nominal center can switch
back and forth across output times.
To avoid ambiguity in assigning subhalo bound and infall mass, our
tracking method now eliminates switches by requiring that once a subhalo
is a central, it remains so until falling into a more massive halo.
This history-based approach assigns as central the oldest subhalo in the host
halo, which would typically correspond to the most massive and evolved galaxy,
better mapping to how central galaxies are defined observationally.
However, since this criterion is based on subhalo history, while
{\sl Subfind\/} only uses the instantaneous density field to define the
central (most massive) subhalo, the two definitions of subhalo centrality
are in conflict in a small fraction cases.
In these cases, we defer to the history-based approach, and we swap the
subhalo bound masses of the history-based central and {\sl Subfind\/}-based
central, such that the history-based central is always the most massive
subhalo, and by extension the most massive galaxy.
We find this produces more stable subhalo catalogs, in terms of individual
mass histories and the resultant subhalo mass function.

\subsection{The Stellar Mass of Subhalos} \label{sec:minf}

Since our goal is to test the correspondence between subhalos and galaxies, a critical issue is how to relate stellar mass to a subhalo.
The simplest prescription assigns stellar mass to a subhalo based on its instantaneous bound dark mass or maximum circular velocity.
However, this method leads to satellite subhalos with significantly less concentrated radial distributions than seen for satellite galaxies around the Milky Way \citep{KraGneKly04} or galaxy clusters \citep{DieMooSta04}.
Better agreement with observations has been found using semi-analytic models of star formation applied to subhalo catalogs \citep{SprWhiTor01,DiaKauBal01}, which correlate stellar mass to a subhalo's mass at the time the stellar component was assembled (before infall), thus indicating that it necessary to use subhalo mass histories to accurately assign stellar mass \citep{GaoDeLWhi04}.

The motivation for this approach comes from the following physical model.
As a halo falls into a larger one and becomes a subhalo, tidal stripping quickly removes bound mass from the outer regions of the subhalo, giving rise to significant mass loss shortly after accretion.
This behavior has been explored both in simulations  \citep[e.g.,][]{DieKuhMad07,LimSomNat09,PenNavMcC08,WetCohWhi09a} and observationally through galaxy-galaxy lensing \citep{LimKneBar07,NatDeLSpr07,NatKneSma09}.
However, the galaxy, being more compact and at the center of the subhalo, remains intact longer.
Thus, dark mass stripping does not directly correspond to stellar mass stripping.
The subhalo's gaseous component -- being more diffuse and readily affected by ram-pressure stripping, tidal shocks, and adiabatic heating in the hot group/cluster environment -- quickly becomes stripped after infall, halting subsequent star formation in the satellite galaxy.
Thus, galaxy stellar mass is expected to correlate with the subhalo's mass immediately before infall \citep{NagKra05}.

Observationally, there are a variety of claims for the star formation efficiency of satellite galaxies, with significant systematic dependence on group/cluster definition or HOD parametrization, criteria for blue/red (star forming/quenched) split, and inclination effects \citep[see][for a comparison of HOD color models]{Ski09}.
The HOD analysis of the Sloan Digital Sky Survey (SDSS) of \citet{ZehZheWei05} implies few faint blue satellites, and that at higher luminosity blue satellites are $3-5$ times less common than red satellites in groups.
Examining group catalogs from SDSS, \citet{WeiKauvdB09} find an appreciable fraction ($>20\%$) of satellites in groups are blue, and fit a simple fading
model where star formation in satellites quenches $\sim2$~Gyr after infall \citep[see also][]{KanvdB08}.
Using a subhalo catalog with an analytic prescription for star formation and quenching, \citet{WanLiKau07} find an exponential decay time for satellite star formation after infall of $\sim2.3$~Gyr.
At $z\sim1$, the clustering results of \citet{CoiNewCro08} imply that
immediate star formation quenching at infall is likely too restrictive
at higher redshift.

Hydrodynamic simulations also indicate short yet finite timescales for halting star formation.
High-resolution SPH simulations at $z\sim0$ indicate that satellite star formation can continue for up to 1 Gyr after infall \citep{SimWeiDav09}.
While the assumption that satellite gas accretion rates are significantly lower than those of central galaxies may break down at $z>1$ \citep{KerKatFar09}, the significantly more recent infall times of satellites at high redshift mediates this concern. 
Promisingly, \citet{NagKra05} find good agreement in comparing the radial profiles of subhalos in dark matter-only simulations selected on infall mass with galaxies selected on stellar mass in dark matter plus hydrodynamic simulations.

Despite the above uncertainties, attempts to use subhalo infall mass (or infall maximum circular velocity) to assign stellar mass have been successful at reproducing various observed galaxy properties at a range of redshifts \citep{BerBulBar06,ConWecKra06,ValOst06,WanLiKau06,YanMovdB09}, and so this is the approach we use here.
The success of this method may be a derivative of its implicit allowance for some star formation after infall, as we describe below.
Furthermore, our analysis is restricted to $z<1$, where the approximation of rapid quenching of star formation is most valid.

\subsection{Assigning Infall Mass and Stellar Mass} \label{sec:mstel}

\begin{figure}
\begin{center}
\resizebox{3.3in}{!}{\includegraphics{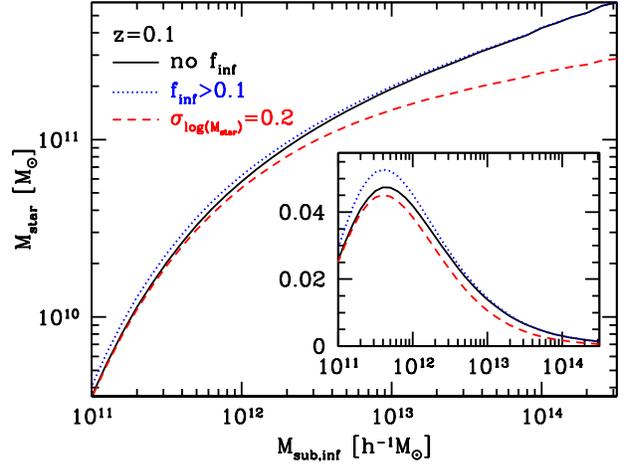}}
\end{center}
\vspace{-0.1in}
\caption{
Relation between stellar mass, $M_{\rm star}$, and subhalo infall mass, $M_{\rm sub,inf}$, using subhalo abundance matching against the observed stellar mass function of \citet{ColNorBau01}.
Solid curve shows relation to our full subhalo catalog (no removal threshold), dotted curve shows the relation under our maximal threshold for satellite removal, and dashed curve shows the mean relation when using a log-normal $0.2$ dex scatter in $M_{\rm star}$ at a fixed $M_{\rm sub,inf}$.
Inset shows ratio $M_{\rm star}/M_{\rm sub,inf}$.
} \label{fig:mstel}
\end{figure} 

Motivated by its expected correlation with stellar mass, our scheme for assigning subhalo infall mass, $M_{\rm inf}$, is as follows.
Centrals subhalos are assigned $M_{\rm inf}$ as their current bound mass.
This mass represents halo mass not bound to any satellite subhalos and so it should track the gas mass available to accrete onto the central galaxy.
We assign to a satellite subhalo its mass when it fell into its current host halo (its subhalo mass when it was last a central subhalo).
With the expectation that a subhalo merger corresponds to a galaxy merger, if two satellite subhalos merge, their satellite child is given the sum of 
their infall masses.
Thus, our tracking scheme inherently incorporates stellar mass growth though satellite-satellite mergers, an often-ignored but non-trivial contribution to galaxy evolution \citep{WetCohWhi09a,WetCohWhi09b,KimBauCol09}.
If a satellite subhalo's orbit brings it outside its host halo, thus becoming its own distinct smaller halo, the subhalo is assigned its satellite parent's infall mass during the output immediately following ejection, but is assigned its current bound mass for subsequent outputs if it remains a separate central subhalo.

We assign stellar mass to our subhalo catalog using subhalo abundance matching (SHAM) such that by construction we recover the observed stellar mass function.
We rank order our subhalo catalog by infall mass, and assign stellar mass such that $n(>M_{\rm inf})=n(>M_{\rm star})$, using the observed stellar mass function of \citet{ColNorBau01}.
Figure~\ref{fig:mstel} shows the $M_{\rm star}-M_{\rm inf}$ relation, for our full satellite subhalo catalog (no removal threshold) and for the maximal threshold for satellite removal we consider.
Since the satellite fraction increases at lower mass, the discrepancy is strongest there, though the offset is comparable to the error in the observed stellar mass function \citep{ColNorBau01}.
The star formation efficiencies in Fig.~\ref{fig:mstel} are in good agreement with the constrains from weak lensing of \citet{ManSelKau06} and also agree with the Milky Way, which has stellar mass of $5\times10^{10}M_\odot$ and a dark halo mass of about $2\times10^{12}M_\odot$ \citep{BinTre}, making $M_{\rm star}/M_{\rm halo} \approx 3\%$.
See \citet{MosSomMau09} for a detailed comparison of derived $M_{\rm star}-M_{\rm subhalo}$ relations in the literature.

Despite the seeming requirement for rapid satellite quenching after infall, SHAM does implicitly allow for satellite star formation.
This is because we match stellar mass to satellites' infall mass at an observational epoch, not at the time of infall.
At $z<1$, satellite times since infall are typically several Gyr, and during that time satellite $M_{\rm inf}$ remains fixed.
Thus, if there is appreciable evolution (growth) in the global $M_{\rm star}-M_{\rm inf}$ relation during that time, this will manifest itself as stellar mass growth for satellites.
We have explored additional satellite star formation after infall by increasing all satellite infall masses by a fixed fraction, but we find that this is simply degenerate with increasing our satellite removal threshold given in the next section.

While the SHAM technique assumes a one-to-one correspondence between $M_{\rm inf}$ and $M_{\rm star}$, there may be considerable scatter in the relation.
Fits to the spatial clustering of luminosity or stellar mass limited galaxies samples frequently require a ``soft'' turn on in the central galaxy
occupation least they overpredict the large-scale clustering, which would naturally arise from scatter in observable at fixed (sub)halo mass.
For example, \citet{ZheCoiZeh07}, applying Halo Occupation Distribution (HOD) modeling to galaxies at $z\sim0$ and $\sim1$, find log-normal scatter in luminosity at fixed halo mass of $0.15-0.3$ dex, decreasing with increasing halo mass, consistent with the scatter of $\sim0.15$ dex found in the group catalogs of \citet{YanMovdB08a} at $z\sim0$.
Similarly, \citet{vdBYanMo07} find a lower limit of $0.2$ dex scatter in halo
mass at fixed luminosity from conditional luminosity function modeling,
consistent with the results of \citet{MorvdBCac09} from satellite kinematics.
\citet{TasKraWec04} find good agreement in spatial clustering when matching subhalos to galaxies if they imposed significant scatter ($0.6$ dex) in luminosity at fixed subhalo maximum circular velocity.

To examine the importance of this scatter when comparing against observed
galaxy samples, we also produce a stellar mass catalog by imposing a fixed
$0.2$ dex (log-normal) scatter in stellar mass at fixed subhalo infall mass
such that we recover the observed SMF.
The mean $M_{\rm star}$ at a fixed $M_{\rm inf}$ is shown as the dashed
curve of Fig.~\ref{fig:mstel}.
Since the mass function steeply falls with mass, introducing scatter biases
$M_{\rm star}$ to a lower value at a fixed $M_{\rm inf}$.
Higher scatter also causes a stellar mass threshold to correspond to a lower effective subhalo $M_{\rm inf}$ threshold.\footnote{
This is not immediately apparent in Fig.~\ref{fig:mstel}, where it might appear that introducing scatter causes \textit{increased} effective $M_{\rm inf}$ at fixed $M_{\rm star}$.
This is because we show the mean $M_{\rm star}$ at a given $M_{\rm inf}$, which is qualitatively different than the mean $M_{\rm inf}$ at a given $M_{\rm star}$.}
This effect is particularly strong above $M_{\rm inf} \approx 10^{12}\,h^{-1}M_\odot$, where the subhalo mass function starts to fall exponentially with mass.

We follow a similar approach to SHAM when we compare to observations of 
magnitude limited samples (\S\ref{sec:obscompare}), matching the number 
densities as $n(>M_{\rm inf})=n(>L)$.
The two assumptions -- that stellar mass and luminosity are set by infall 
mass -- are not fully consistent since luminosity may evolve even if stellar mass does not.
However, it is not unnatural to assume the rank ordering is preserved in a population sense.

\section{Impact of Satellite Removal} \label{sec:thresholds}

As a subhalo orbits in its host halo, dynamical friction removes energy from its orbit, bringing it closer to the central galaxy, while tidal forces strip its mass.
While the satellite galaxy is expected to retain its stellar mass throughout most of the subhalo's mass stripping, eventually (though possibly in longer than a Hubble time) the galaxy will become removed from the satellite population, either from merging with the central galaxy or becoming disrupted into the diffuse ICL (or both).
We now examine the impact different removal criteria have on the satellite population.
We first focus on physically informative properties in the context of the
halo model \citep{Sel00,PeaSmi00,BerWei02,CooShe02}, including the HOD and radial profile.
However, since these are not direct observables, we reserve detailed
comparisons with observations, via spatial clustering and satellite fractions,
to the next section.

As described in the introduction, various prescriptions for satellite subhalo
removal have been used.
We focus primarily on the one used most often in previous work, which is also the most straightforward given our tracking: remove subhalos where the instantaneous bound mass to infall mass ratio, $f_{\rm inf}= M_{\rm bound}/M_{\rm inf}$, falls below some threshold.
While dark matter-only simulations do not incorporate the complex hydrodynamics involved in galaxy evolution, if a satellite galaxy's gas is stripped before its stellar component, then it is conceivable that a simple criterion on subhalo dark matter stripping yields a good approximation for galaxy stellar mass removal since both experience purely gravitational interactions.
We examine a range in threshold $f_{\rm inf}$ from $0.01-0.1$, which as we show in \S\ref{sec:obscompare} brackets the observations, and even subhalos stripped to $f_{\rm inf}=0.01$ remain well-resolved in the mass ranges we consider below.
We refer to our subhalo catalog with no removal threshold as the ``full'' tracking model.
Based on convergence tests, our subhalo catalog at $z=0.1$ is robust to artificial disruption down to $\sim10^{11.5}\,h^{-1}M_\odot$, so the full tracking model serves as an upper limit to the satellite population.
We explore different removal criteria in \S\ref{sec:comparison}.

\subsection{Halo Occupation Distribution} \label{sec:hod}

\begin{figure}
\begin{center}
\resizebox{3.3in}{!}{\includegraphics{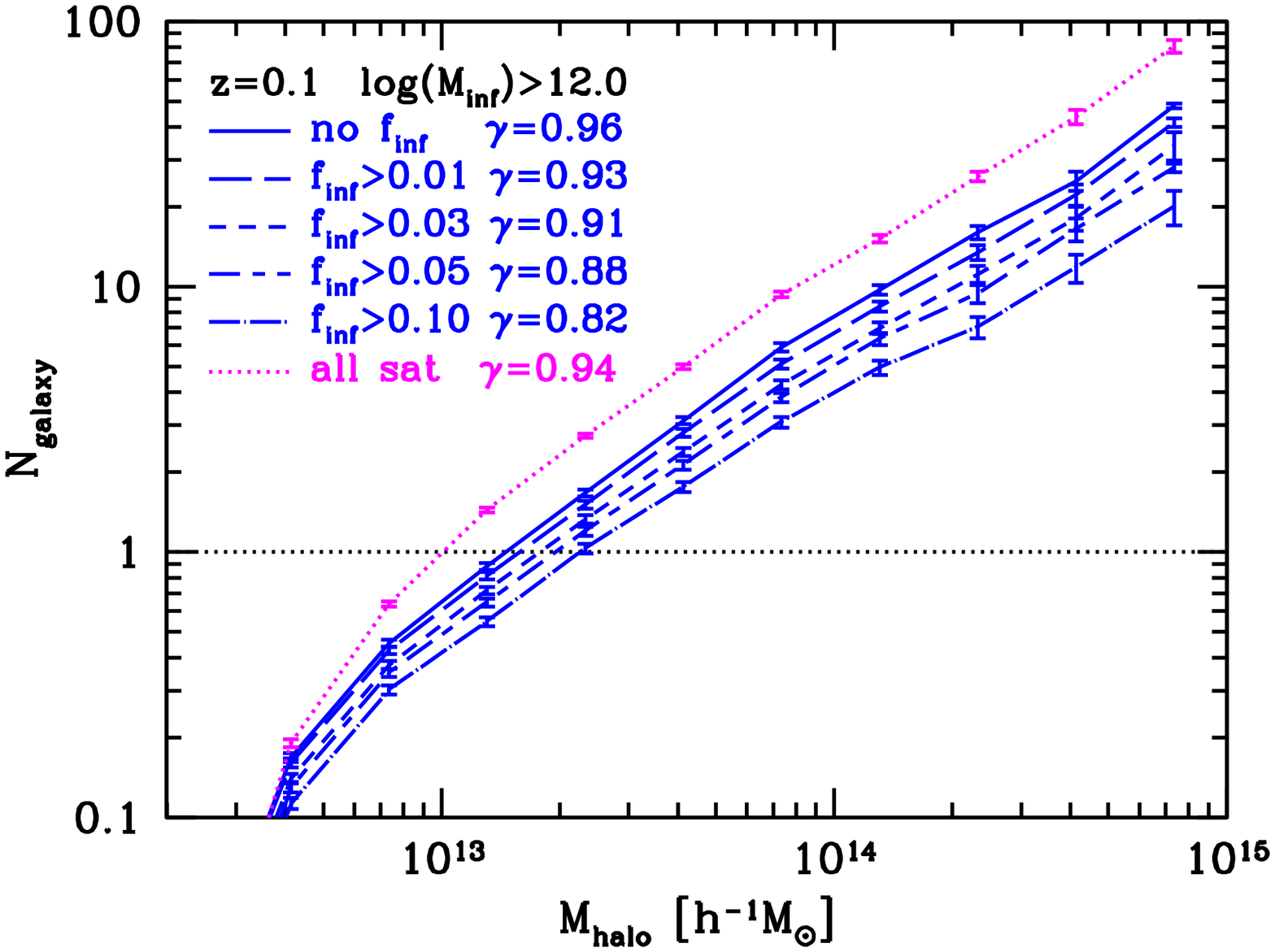}}
\resizebox{3.3in}{!}{\includegraphics{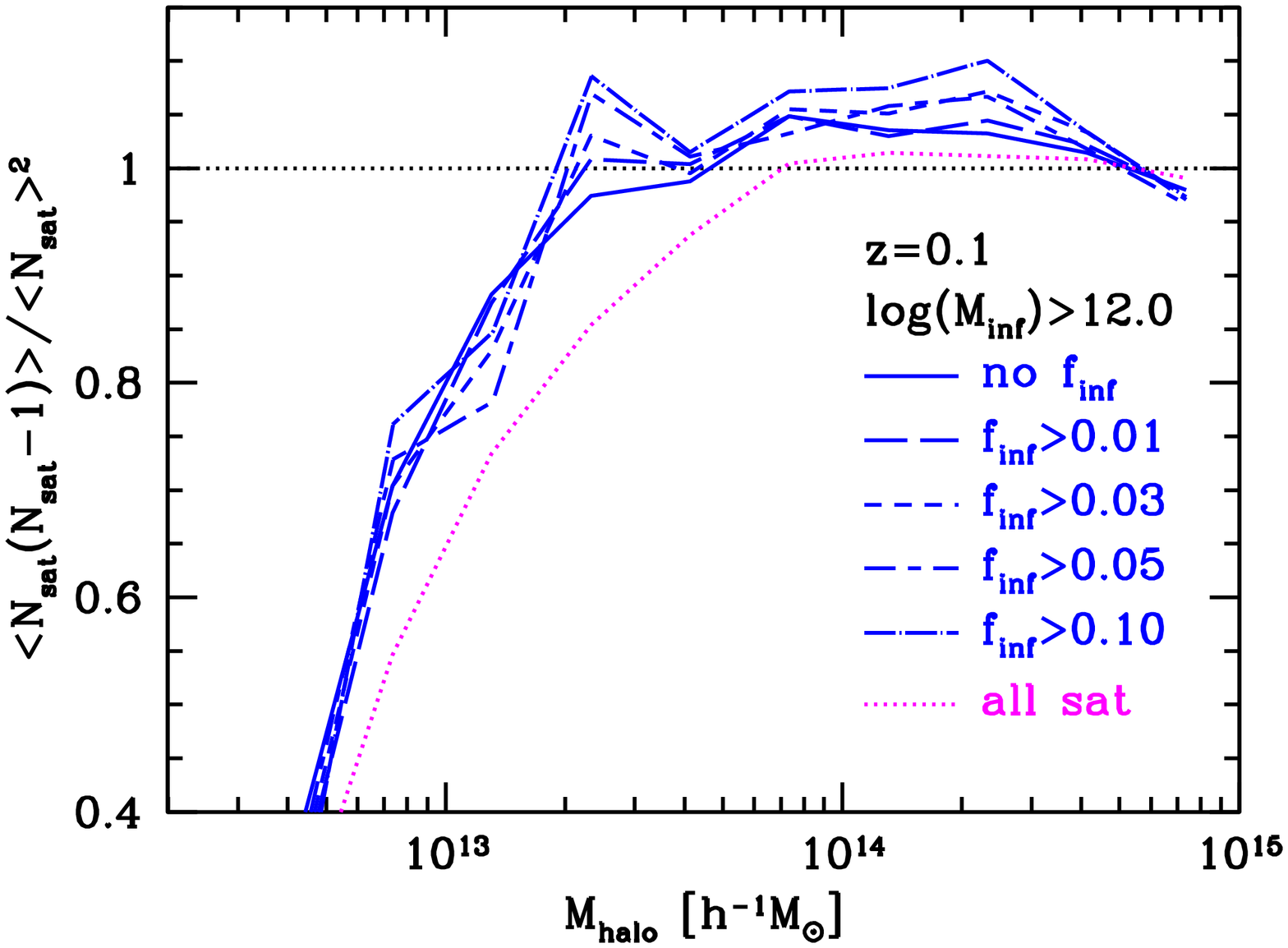}}
\end{center}
\vspace{-0.1in}
\caption{
\textbf{Top}: Halo Occupation Distribution (HOD) for satellite subhalos with $M_{\rm inf} >10^{12}\,h^{-1}M_\odot$ ($M_{\rm star}>10^{10.8}\,M_\odot$, $L>0.6\,L_*$), for different minimum removal thresholds, $f_{\rm inf}=M_{\rm bound}/M_{\rm inf}$.
Solid curve shows all resolved satellites, regardless of mass stripping (``no $f_{\rm inf}$''), while dotted curve shows all satellites ever accreted, even if they fall below the resolution limit (a model in which satellites never merge or disrupt).
$\gamma$ indicates best-fit slope to Eq.~\ref{eq:hod}.
Errors are standard deviation of the mean HOD at fixed halo mass.
\textbf{Bottom}: Normalized second moment of satellite HOD.
} \label{fig:hod}
\end{figure}

Figure~\ref{fig:hod} (top) shows the Halo Occupation Distribution (HOD) for our full tracking model (``no $f_{\rm inf}$'') and for different minimum $f_{\rm inf}=M_{\rm bound}/M_{\rm inf}$ thresholds.
We fit the HOD to
\begin{equation} \label{eq:hod}
  N_{\rm sat} = \left(\frac{M_{\rm halo}}{M_1}\right)^\gamma
  e^{-M_{\rm cut}/M_{\rm halo}}
\end{equation}
with $M_1$, $M_{\rm cut}$ and $\gamma$ as free parameters.
The slope, $\gamma$, is given in Fig.~\ref{fig:hod} for each value of 
$f_{\rm inf}$.
Raising the threshold for satellite removal, $f_{\rm inf}$, affects satellites more in higher mass halos, leading to a shallower slope.
As the threshold increases, the HOD shoulder also becomes broader, meaning that 
more halos host only one galaxy.
This indicates that satellite subhalos of a given $M_{\rm inf}$ are stripped more in higher mass halos, because of the stronger tidal fields and longer dwell times.
This implies that the details of modeling satellite galaxy removal are more critical in cluster-mass halos than in lower mass groups.

The dotted curve also shows the HOD if satellites are never removed by disruption or merging with the central galaxy, in other words, every infalling galaxy survives until $z=0.1$ (which we measure by analytically retaining every infalling halo).
This provides a strict upper limit to the HOD, contingent only upon the
accuracy of the halo merger trees and not on subhalo tracking and resolution.
As we explore in \S\ref{sec:obscompare}, this scenario is highly disfavored.

In the halo model, the spatial clustering on large scales is dominated by galaxy pairs in separate halos (``2-halo'' term), which depends on the first moment of the HOD, $\langle N \rangle(M_{\rm halo})$, where $N=N_{\rm sat}+1$.
However, the clustering on small scales is dominated by pairs within a halo (``1-halo'' term), which depends on the second moment, $\langle N(N-1) \rangle(M_{\rm halo})$.
Figure~\ref{fig:hod} (bottom) shows the normalized second moment of the satellite HOD, defined as $\langle N_{\rm sat}(N_{\rm sat}-1) \rangle/\langle N_{\rm sat} \rangle^2$.
The normalized second moment asymptotes to $\approx1$ at high halo mass, reflecting a nearly Poisson distribution, and it approaches zero at low halo mass where the probability of hosting one satellite remains finite while the probability of hosting two satellites goes to zero.
Figure~\ref{fig:hod} shows some dependence on removal threshold, particularly at high halo mass where raising the threshold increases the second moment.
This is driven by correlated infall, such as multiple satellites infalling as a group, and thus correlated amounts of mass stripping.
This is supported by the fact that the ``no removal'' scenario (dotted curve), which represents all satellites ever accreted and is insensitive to mass stripping, is closer to Poisson.

\subsection{Satellite Radial Density Profile} \label{sec:profile}

\begin{figure}
\begin{center}
\resizebox{3.3in}{!}{\includegraphics{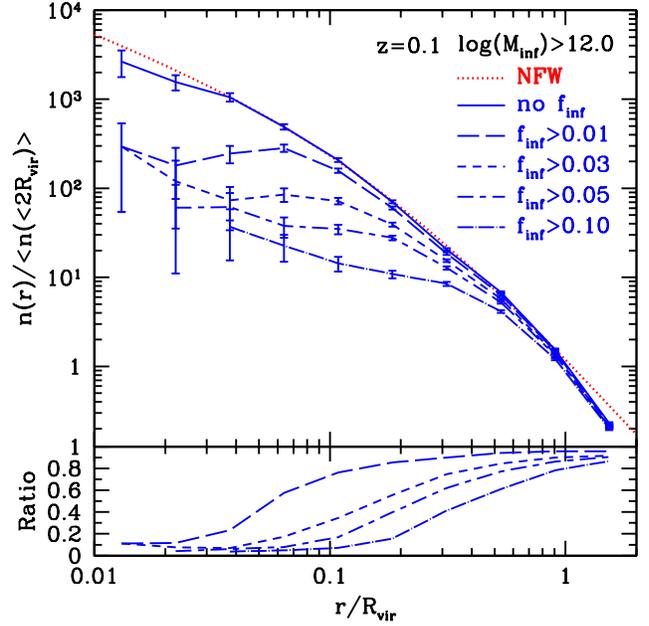}}
\end{center}
\vspace{-0.1in}
\caption{
\textbf{Top}: Satellite radial density profile for satellites with $M_{\rm inf}>10^{12}\,h^{-1}M_\odot$ ($M_{\rm star}>10^{10.8}\,M_\odot$, $L>0.6\,L_*$).
Solid curve shows all resolved satellites (no removal threshold), while subsequent curves show different minimum removal thresholds.
Radii are defined by distance to central galaxy and are scaled to the host halo virial radius, $R_{\rm vir}$.
Densities are normalized to the number within $2\,R_{\rm vir}$ using no threshold.
Errors bars indicate Poisson error in each radial bin.
\textbf{Bottom}: Ratio of profile using a given $f_{\rm inf}$ threshold to that of using no removal threshold.
} \label{fig:profile}
\end{figure}

In addition to the number of satellites in a halo, we also explore how removal
thresholds affect their locations.
Figure~\ref{fig:profile} (top) shows the satellite radial density profile,
obtained from stacking all halos hosting the given satellite subhalos.
Radii are defined by the distance of a satellite to its central galaxy and are scaled to the host halo's virial radius, $R_{\rm vir}$, obtained from the halo FoF mass and concentration assuming a spherical NFW density profile ($R_{\rm vir} \approx R_{200c}$ at $z \approx 0$).
Since halos are triaxial, particularly for those with significant substructure
(recent mergers), the assumption of spherical symmetry is highly approximate
\citep[see e.g.,][Fig.~2]{TreePM}.
Additionally, subhalos preferentially lie in a plane defined by their host halo's major axis \citep[e.g.,][]{KroTheBoi05,FalJinLi08,LibFreCol09}, behavior also found for galaxies in SDSS \citep{Bra05} and dwarf satellites around the Milky Way \citep{MetKroJer09}.
This implies that the assumption of spherical symmetry is even less valid for subhalo populations, and that a spherical overdensity halo finder will miss satellites aligned along the halo major axis.
Because of these effects, we find that selecting satellites out to $R_{\rm vir}$ only captures $70\%$ of the entire satellite population, while going out to $2\,R_{\rm vir}$ captures $95\%$.
Hence, number densities are normalized by the counts (using no removal threshold) within $2\,R_{\rm vir}$.

The full subhalo tracking model (no removal threshold) shows satellite subhalos tracing the NFW profile to good approximation down to $0.01\,R_{\rm vir}$, though this may represent over-resolution.
Since stripping is more prominent in the halo's dense central region,
raising the threshold preferentially removes satellites from the halo center,
leading to a less concentrated profile.
Figure~\ref{fig:profile} (bottom) shows the ratios of the profiles using various $f_{\rm inf}$ values to that of the full tracking model.
In all cases, significant suppression occurs below $0.4\,R_{\rm vir}$.
Interestingly, for no $f_{\rm inf}$ threshold does the profile asymptote to unity, implying that mass stripping affects satellites even at large radii.
The asymptotic value has no significant dependence on $M_{\rm inf}$ or on time since infall, indicating that this effect is not driven by satellites that have passed through the halo center, but instead by subhalos being significantly stripped during or soon after infall.

These profiles can be compared with observationally measured satellite surface density profiles in galaxy groups/clusters, a method typically based on stacking the profiles of galaxies from several groups/clusters.
Some work indicates that satellite galaxies trace well their host halo's NFW profile.
Using a compilation of local galaxy clusters, \citet{DieMooSta04} find that the surface density profiles of their member galaxies match that cluster-mass halo particles in simulations within $20\%$ scatter and no systematic offset down to $0.05\,R_{\rm vir}$.
Similarly, \citet{vdBYanMo05} examine satellite galaxies in Two Degree Field Galaxy Redshift Survey (2dFGRS), finding that they are consistent with following their host halo's density profile, though they were not able to discriminate distributions in halo cores.

Other work indicates that, while satellite density profiles can be fit by an NFW profile, they require a lower concentration parameter than the dark matter.
Using clusters of mass $>3\times10^{13}\,h^{-1}M_\odot$, \citet{LinMohSta04} find that galaxy profiles down to $0.02\,R_{\rm vir}$ require an NFW concentration parameter of $c=2.9 \pm 0.2$.
A similar analysis by \citet{MuzYeeHal07} finds that galaxies trace NFW with $c=4.1 \pm 0.6$ down to $0.5\,R_{\rm vir}$.
Examining lower mass ($<10^{14}\,h^{-1}M_\odot$) galaxy groups/clusters in SDSS, \citet{HanMcKWec05} find that satellite profiles down to $0.1\,R_{\rm vir}$ can be well-fit with NFW profiles using $c<3$, with similar results from the 2dFGRS and SDSS group catalogs of \citep{YanMovdB05b}.

Concentrations of $c=3-4$ are lower than the average halo concentration of $c=5-6$ for cluster-mass halos in our simulations.
Interestingly, an NFW profile with $c=3$ is consistent with the satellite distribution in our simulation down to $0.06\,R_{\rm vir}$ using $f_{\rm inf}=0.01$.
However, differences in halo virial radii estimates, density normalizations, cluster centering, and galaxy luminosity thresholds make a detailed comparison difficult, as highlighted by the diversity in concentrations derived observationally.

One strong systematic uncertainty in measuring the profiles is how one chooses the halo center.
Two common methods of defining a cluster center are the location of the Brightest Cluster Galaxy (BCG) and the peak of X-ray temperature profile.
However, these are often offset by several $10$'s of kpc.
For example, for rich clusters in the SDSS MaxBCG catalog, the median offset between the BCG and X-ray center is $58\,h^{-1}$~kpc \citep{KoeMcKAnn07b}.
To estimate how such uncertainties propagate, we examine the degree to which centering changes the measured density at $0.1\,R_{\rm vir}$ by giving the central subhalo an arbitrary offset.
We find that an offset of $10$, $25$, or $50\,h^{-1}$~kpc leads to a change of $+1\%$, $+4\%$ and $-4\%$ in the measured density, the error growing rapidly with offset amplitude.
Requiring less than a $10\%$ reduction in measured density at $0.1\,R_{\rm vir}$  for clusters requires an offset of less than $60\,h^{-1}$~kpc in the central subhalo.
Such offset may be minimized by selecting BCGs based on their color profiles \citep{BilHoeBab08}.

\subsection{Radius at Removal: Merger vs. Disruption} \label{sec:dist}

\begin{figure}
\begin{center}
\resizebox{3.3in}{!}{\includegraphics{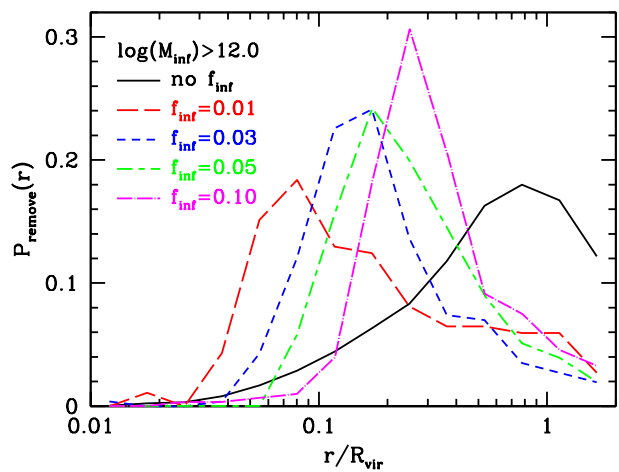}}
\end{center}
\vspace{-0.1in}
\caption{
Satellite radial distribution probability (not divided by volume) at the time of crossing below a given removal threshold, $f_{\rm inf}$, for satellites with $M_{\rm inf}>10^{12}\,h^{-1}M_\odot$ ($M_{\rm star}>10^{10.8}\,M_\odot$, $L>0.6\,L_*$).
For reference, solid curve shows radial distribution of all resolved satellites at $z=0.1$.
} \label{fig:dist}
\end{figure}

While Fig.~\ref{fig:profile} shows the profile of extant satellites, it does not indicate the location of satellites when they became removed and where their central particles, associated with a stellar population, end up.
This would provide a measure of whether satellite galaxies merge with the 
central galaxy or disrupt into the ICL.
One route to examine this would be to track central ``star'' particles in subhalos after they become removed and examine their overall distribution at $z\sim0$ \citep{MurArnGer04,WilGovWad04,SomRomPor05,RudMihMcB06}.
We take a different approach and examine galaxy radial distribution at the time of removal.
While Fig.~\ref{fig:profile} shows satellites \textit{above} a threshold $f_{\rm inf}$, we now examine satellites \textit{as} they fall below the threshold by selecting those within $25\%$ of a given $f_{\rm inf}$ value.

Figure~\ref{fig:dist} shows the probability distribution that a satellite is at a given scaled radius for a given $f_{\rm inf}$ value.
For reference, the solid curve shows the radial distribution of all resolved satellites at $z=0.1$, regardless of stripping.
These are obtained from stacking all halos hosting the given satellites, though the results do not change significantly if we look only at cluster-mass halos.
Satellites fall below a lower $f_{\rm inf}$ threshold at a smaller radius.
This is expected in a model where subhaloes are stripped of mass as dynamical friction brings their orbits to smaller radii.
All thresholds display a broad distribution with radius, since stripping will also depend on satellite orbital parameters and time since infall \citep{BoyMa07}.
However, the distribution is more peaked for higher $f_{\rm inf}$ thresholds.
This likely arises because significantly stripped subhalos are closer to halo centers, where their velocities are both higher and more radial.
Indeed, as the removal threshold is raised from $f_{\rm inf}=0.01$ to $0.1$, the average (absolute) ratio of radial to tangential velocities for the satellites drops from $1.5$ to $1.1$.
The average for the entire satellite population is between these at $1.4$.

To understand their fates, one must know the direction in which satellites are moving as they are removed.
If a satellite disrupts at large radius but is moving toward halo center on a highly radial orbit, its stellar mass may still funnel to the central galaxy.
As a measure of this, we examine the fraction of satellites that are moving net inward towards halo center as they cross below $f_{\rm inf}$.
We find only mild dependence on $f_{\rm inf}$: as the removal threshold is 
raised from $f_{\rm inf}=0.01$ to $0.1$, the fraction of satellites moving inward
falls from $58\%$ to $52\%$.
Both are lower than the average for the entire satellite population of $62\%$.

The above results imply that more highly stripped satellites are more likely moving inward and their orbits are more radial.
However, both of these trends with mass stripping are mild.
Fig.~\ref{fig:dist} shows that, even in our most conservative case of $f_{\rm inf}=0.01$, the (broad) peak of the radial probability distribution occurs at $\sim50\,h^{-1}$~kpc.
This, coupled with the fact that almost half of satellites at removal are moving outward and have significant tangential velocity components, implies that a significant fraction of satellite galaxies disrupt into the ICL, and do not immediately merge with the central galaxy \citep[see also][]{MonMurBor06,ConWecKra07,PurBulZen07,WhiZheBro07}.

\subsection{Analytical Model for Satellite Removal} \label{sec:dynfric}

As an alternative to numerically tracking satellite subhalos, we can assign infalling halos a removal time at accretion and examine what impact different timescales have on the HOD.
While individual merging/disruption times depend on satellite orbital parameters \citep{BinTre,BoyMaQua08,JiaJinFal08}, we use a simpler parametrization of satellite removal time which depends only on the halos' mass ratio at infall, assumed to hold for an ensemble average of satellites.
Here, an infalling satellite halo with mass $M_{\rm sat,inf}$ merges with the central galaxy or becomes disrupted on a timescale
\begin{equation} \label{eq:dynfric}
  t_{\rm dyn} = C_{\rm dyn}
  \ \frac{M_{\rm halo}/M_{\rm sat,inf}}{\ln(1+M_{\rm halo}/M_{\rm sat,inf})}
  \ t_{\rm Hubble}
\end{equation}
where $t_{\rm Hubble} = H^{-1}(z)$, $M_{\rm halo}$ is the larger halo mass, and we leave $C_{\rm dyn}$ as a free parameter to match to different satellite removal thresholds.

\begin{figure}
\begin{center}
\resizebox{3.3in}{!}{\includegraphics{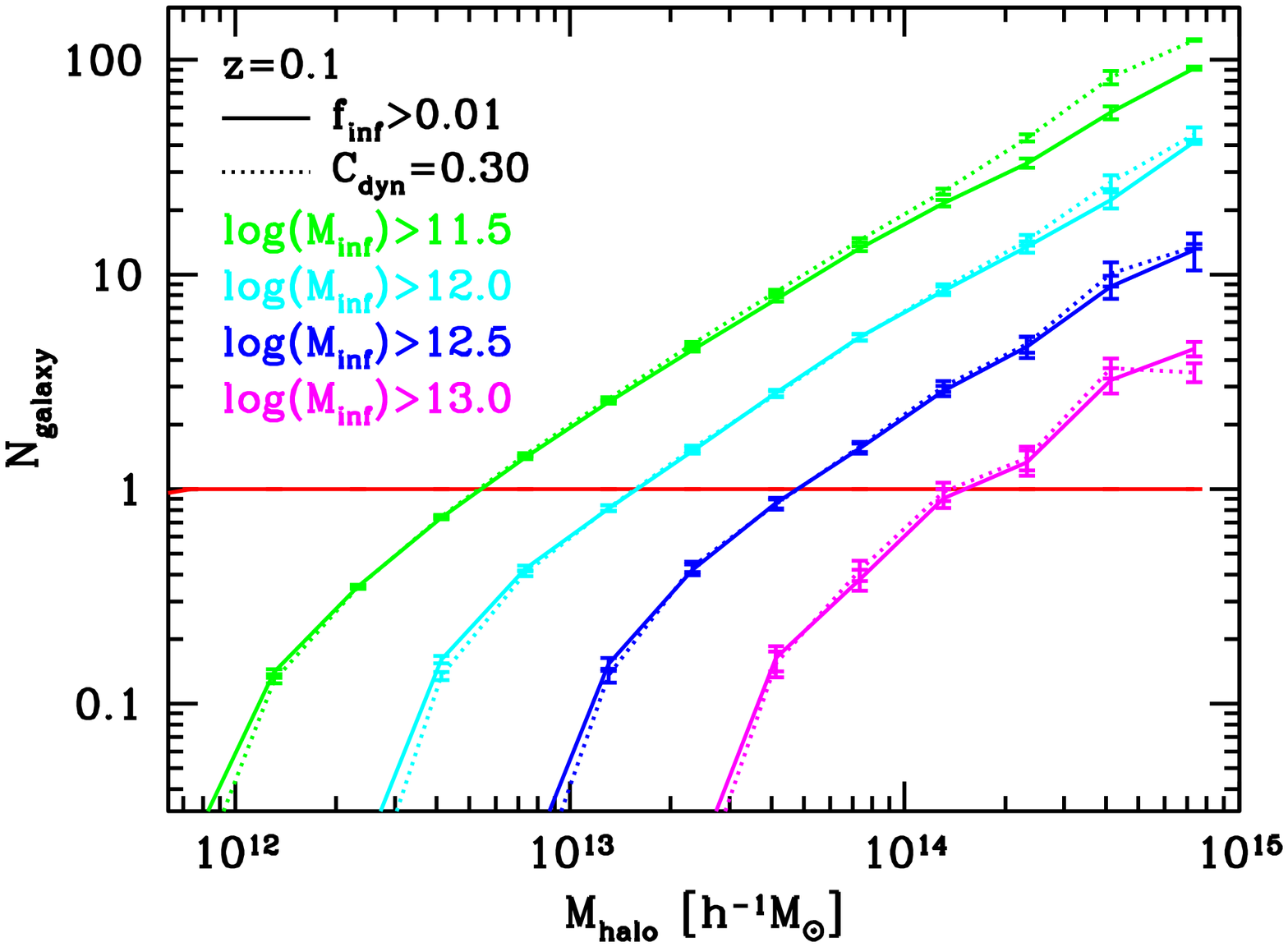}}
\resizebox{3.3in}{!}{\includegraphics{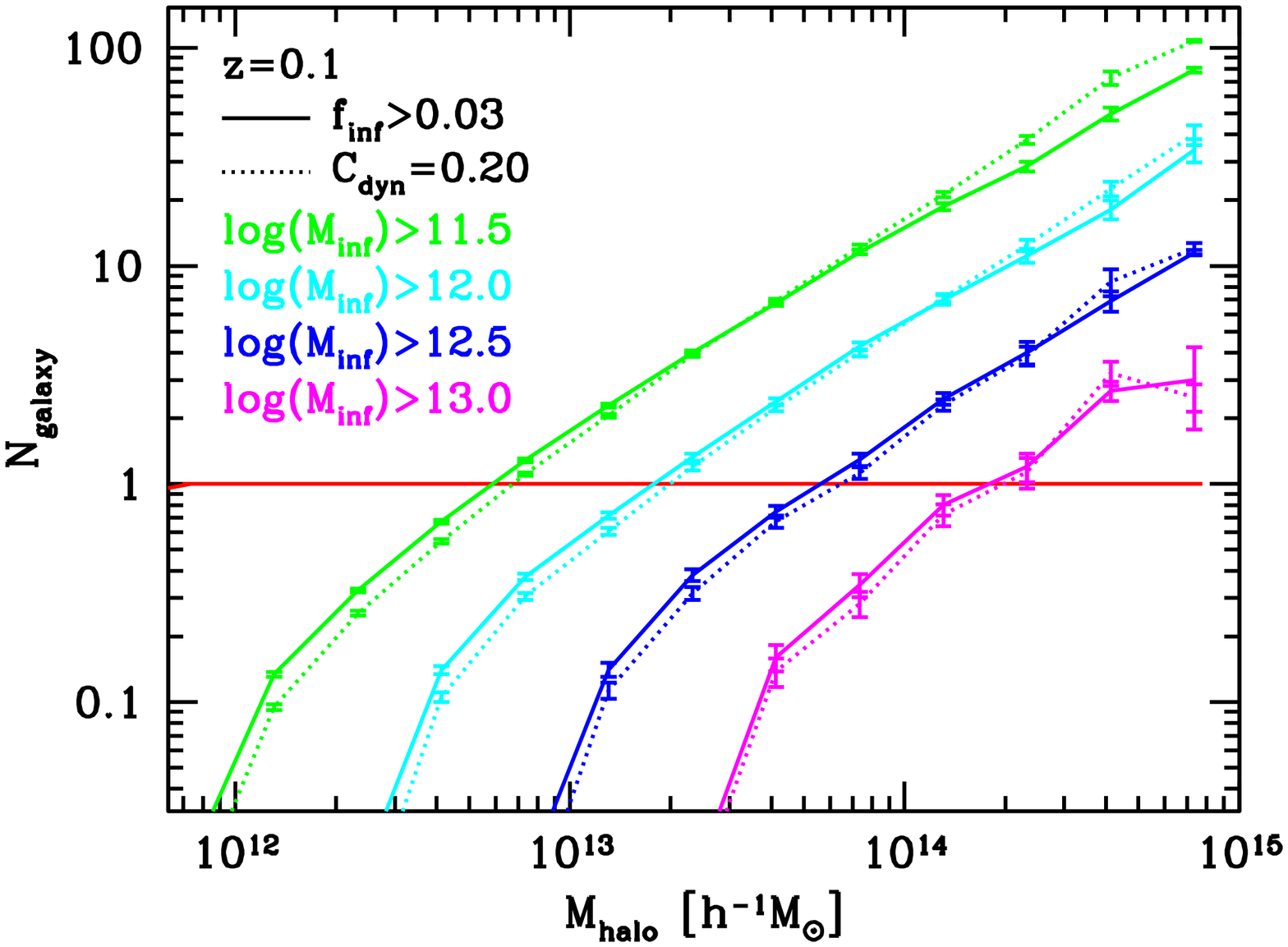}}
\end{center}
\vspace{-0.1in}
\caption{
Halo Occupation Distributions (HOD) for multiple $M_{\rm inf}$-limited samples, using removal thresholds $f_{\rm inf}>0.01$ (top) and $f_{\rm inf}>0.03$ (bottom).
$M_{\rm inf}=10^{11.5},10^{12.0},10^{12.5},10^{13.0}\,h^{-1}M_\odot$ corresponds to $M_{\rm star}=10^{10.3},10^{10.8},10^{11.1},10^{11.3}$ ($L=0.2,0.6,1.3,2.1\,L_*$).
Also shown are HODs for different values of $C_{\rm dyn}$ for dynamical friction infall time, chosen to match the subhalo catalog.
} \label{fig:hodtdyn}
\end{figure}

Figure~\ref{fig:hodtdyn} shows the HOD for multiple $M_{\rm inf}$-limited samples and two removal thresholds, $f_{\rm inf}$.
Also plotted are the HODs using different values of $C_{\rm dyn}$ for dynamical friction.
In each case, the $C_{\rm dyn}$ value is chosen to best match the subhalo catalog using the given $f_{\rm inf}$ value.
Impressively, this sole free parameter can match well the subhalo catalogs across a broad range of subhalo and halo masses, and for multiple removal thresholds.
Additionally, since satellite removal times depend on their orbital parameters the relative robustness of $C_{\rm dyn}$ across a wide halo and subhalo mass range indicates that satellite orbits do not vary significantly with mass, though Fig.~\ref{fig:hodtdyn} shows some hint of shorter infall times (more radial orbits) for satellites in more massive halos.
As we shall see in \S\ref{sec:highz}, using Eq.~\ref{eq:dynfric} with the
same value of $C_{\rm dyn}$ for the given $f_{\rm inf}$ also works well at higher redshift.

\section{Comparisons with Observations} \label{sec:obscompare}

Having examined trends in subhalo properties with varying removal threshold, we now seek to constrain this freedom by comparing with observations.
While the HOD and radial profile of the above sections are physically informative, they are not direct observables.
The most direct comparison with data is via spatial clustering, which provides a scale-dependent test of our subhalo catalog at various masses.
In addition, we compare to observed satellite fractions and cluster satellite luminosity functions, though these measures are less robust since they are derived from HOD modeling of spatial clustering or from constructing group catalogs, both of which are subject to systematic uncertainties in halo mass definition, HOD parametrization, and assumed cosmology.

We again use subhalo abundance matching, as outlined \S\ref{sec:mstel}, but while we matched to the 2dF stellar mass function in \S\ref{sec:mstel}, in this section we match to the number densities of $r$-band luminosity threshold samples in SDSS in order to provide a robust comparison against observed clustering results.
Thus, here the requirements on matching are less restrictive, in that we need only match to a single threshold at a time rather than reproduce the entire luminosity function.

\subsection{Spatial Clustering} \label{sec:wp}

For spatial clustering measurements, we compare against the SDSS galaxy
clustering results of \citet{ZehZheWei05} at $z \approx 0.1$.
To measure spatial clustering, we use the two-dimensional projected galaxy
auto-correlation function
\begin{equation}
  w_p(r_p) = \int_{-\pi_{\rm max}}^{\pi_{\rm max}}d\pi\,\xi(r_p,\pi)
\end{equation}
which has the advantage of effectively integrating over redshift-space distortions, which we incorporate using subhalo velocities.
We use $\pi_{\rm max} = 40\,h^{-1}$~Mpc to match the measurements against which we compare.
This value represents a good balance between integrating over redshift-space
distortions and minimizing correlated noise from large-scale power
\citep{PadWhiEis07}.

Both large- and small-scale clustering are sensitive to the satellite
population.
In the limit that all galaxies are centrals (a halo hosts only one galaxy), 
the galaxy correlation function is simply that of the mass-limited halo sample of the same number density.
As the threshold for satellite removal decreases and the satellite population increases, high-mass halos will host multiple galaxies.
This serves to raise the minimum halo mass of the sample (at fixed number
density) while further weighting the clustering from high-mass halos which
host more satellites.
Satellites in a halo are also close spatially, to each other and to the
central.
Thus, satellites boost both the large- and small-scale clustering.

The small-scale clustering is also dependent on the radial profile of the
satellites, though this sensitivity is mild.
In the limit that central-satellite pairs dominate, the clustering on scale $r_p$ comes from satellites at a large fraction of $r_{\rm vir}$ in small halos and satellites nearer the center of larger halos.
Given the steepness of the mass function and the available halo volume, the
first contribution dominates, weakening the sensitivity to the inner profile.
Therefore, $w_p(r_p)$ is only mildly sensitive to the radial dependence of
satellite removal, though in the extreme limit of no satellites it drops off rapidly on small scales because of halo exclusion.

\begin{figure}
\begin{center}
\resizebox{3.3in}{!}{\includegraphics{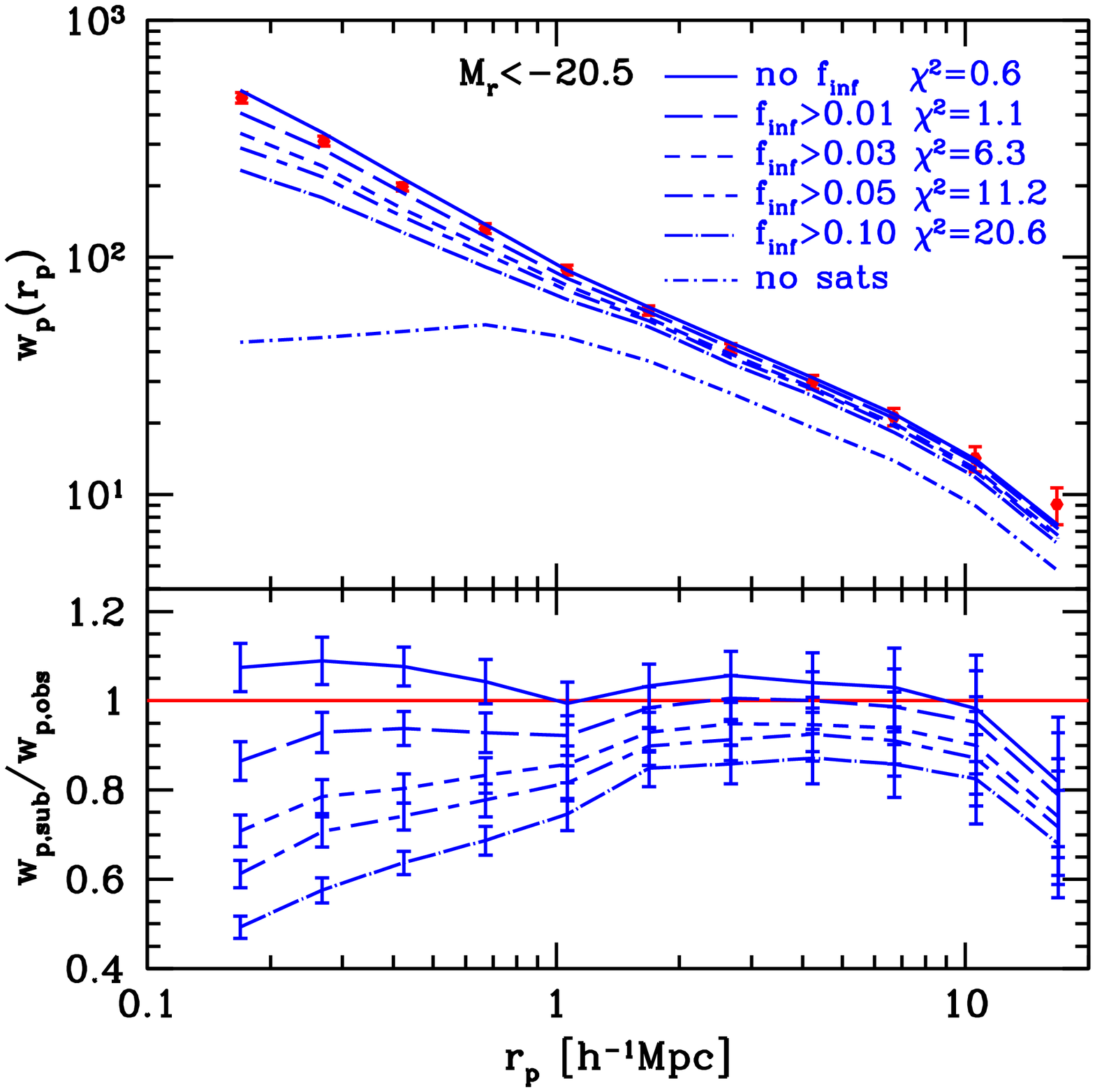}}
\resizebox{3.3in}{!}{\includegraphics{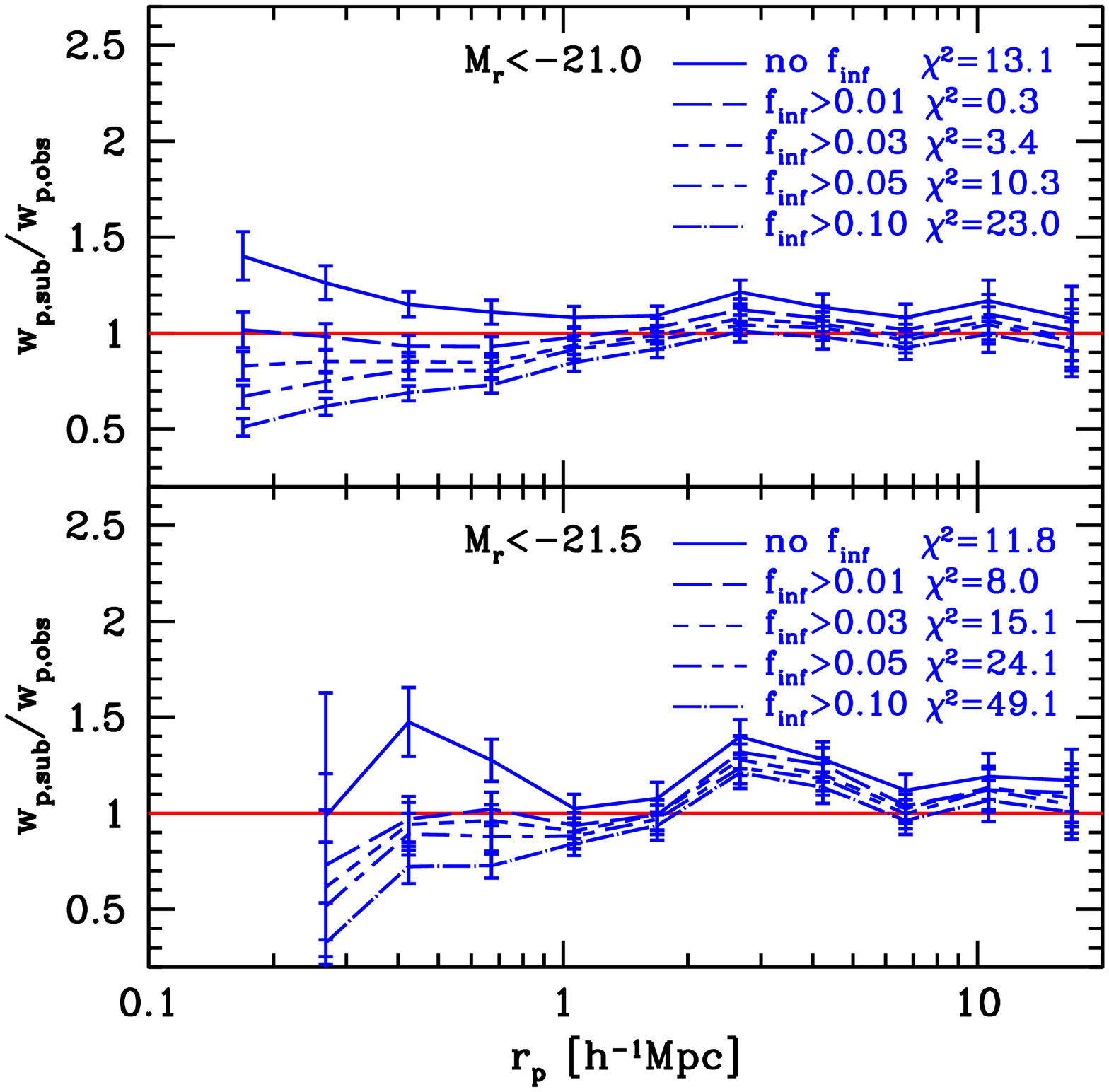}}
\end{center}
\vspace{-0.1in}
\caption{
Projected auto-correlation function at $z=0.1$ for several removal thresholds, $f_{\rm inf}$, (curves) each matched in number density to magnitude-limited samples in SDSS assuming no $L_r-M_{\rm inf}$ scatter and compared with measured $w_p(r_p)$ of \citet{ZehZheWei05} (points).
\textbf{Top}: $w_p(r_p)$ (top) and ratio of simulation to observed $w_p(r_p)$ (bottom) for $M_r<-20.5$ sample.
\textbf{Bottom}: Ratio of simulation to observed $w_p(r_p)$ for $M_r<-21.0$ and $<-21.5$ samples.
$M_r<-20.5$, $-21.0$, and $-21.5$ correspond to subhalo $M_{\rm inf}\gtrsim10^{12}$, $10^{12.5}$, and $10^{13}\,h^{-1}M_\odot$.
Also shown is the reduced $\chi^2$ of the fit to observation for each $f_{\rm inf}$.
} \label{fig:wp}
\end{figure}

Figure~\ref{fig:wp} shows $w_p(r_p)$ of luminosity-threshold samples from SDSS.
These are compared against the $M_{\rm inf}$-threshold subhalo catalog for several removal thresholds, $f_{\rm inf}$, each matched to the threshold number density of the SDSS sample.
The magnitude-limited samples of $M_r<-20.5$ and $<-21.5$ correspond to subhalo $M_{\rm inf}$-limited thresholds of $M_{\rm inf}\approx 10^{12}$ and $10^{13}\,h^{-1}M_\odot$, spanning a range we are able to probe with robust resolution and good statistics.
Impressively, the simple prescription for satellite removal of $f_{\rm inf}=0.01-0.03$, combined with simple number density matching to assign luminosity, matches well at small and large scales, across over an order of magnitude in $M_{\rm inf}$.
We find similar agreement with the $M_r<-20.0$ sample, corresponding to $M_{\rm inf}>10^{11.75}\,h^{-1}M_\odot$, if the ``Sloan Great Wall'' \citep{GotJurSch05} is removed, but because of the involved uncertainty we do not include this sample to constrain $f_{\rm inf}$.
The $M_r<-19.5$ sample, corresponding to $M_{\rm inf}>10^{11.5}\,h^{-1}M_\odot$, agrees as well but is on the edge of robust resolution, so it does not provide a robust constraint.

\begin{figure}
\begin{center}
\resizebox{3.3in}{!}{\includegraphics{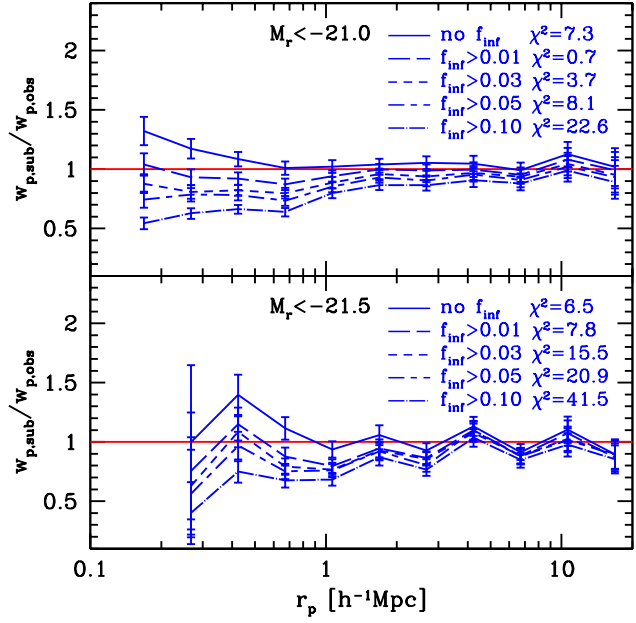}}
\end{center}
\vspace{-0.1in}
\caption{
Same as Fig.~\ref{fig:wp} (bottom), but luminosity is matched to subhalos assuming $0.2$ dex scatter in $L_r$ at fixed $M_{\rm inf}$.
} \label{fig:wpscat}
\end{figure}

We also investigate the influence of luminosity-mass scatter.
Because of the steepness in the subhalo mass function, adding scatter lowers the effective $M_{\rm inf}$ threshold at a fixed number density, and this effect is stronger at higher $M_{\rm inf}$ (see Fig.~\ref{fig:mstel}).
Thus, Fig.~\ref{fig:wpscat} shows $w_p(r_p)$ for most luminous samples, assuming $0.2$ dex scatter in luminosity at fixed $M_{\rm inf}$.
As expected from the decreased effective mass threshold, adding scatter reduces the clustering amplitude, and the effect is stronger at larger scales.
However, the goodness of agreement between observations and $f_{\rm inf}=0.01-0.03$ is not significantly changed, except for the $M_r<-21.5$ sample where agreement is improved somewhat, suggesting necessary scatter at high mass.

\subsection{Satellite Fraction} \label{sec:satfrac}

We next compare against the observed satellite fraction: the fraction of galaxies above a given luminosity threshold that are satellites.
Since the satellite fraction is an integral over the HOD, it integrates the effects of \S\ref{sec:thresholds} into a single number as a function of $M_{\rm inf}$: as the threshold for removal is raised, the satellite fraction drops.

We compare against a variety of values derived from SDSS and 2dFGRS, all at a median $z=0.05-0.1$.
Again we stress that determining satellite fractions from observations requires significant modeling. 
\citet{ZehZheWei05} and \citet{ZheCoiZeh07} satellite fractions were both derived from fitting HODs to the number density and spatial clustering of galaxy samples from SDSS, though they used different HOD parametrizations.
\citet{TinNorWei07} used a similar technique applied to 2dFGRS.
Additionally, \citet{YanMovdB08a} satellite fractions were obtained directly from the SDSS group catalogs of \citet{YanMovdB07}.

\begin{figure}
\begin{center}
\resizebox{3.3in}{!}{\includegraphics{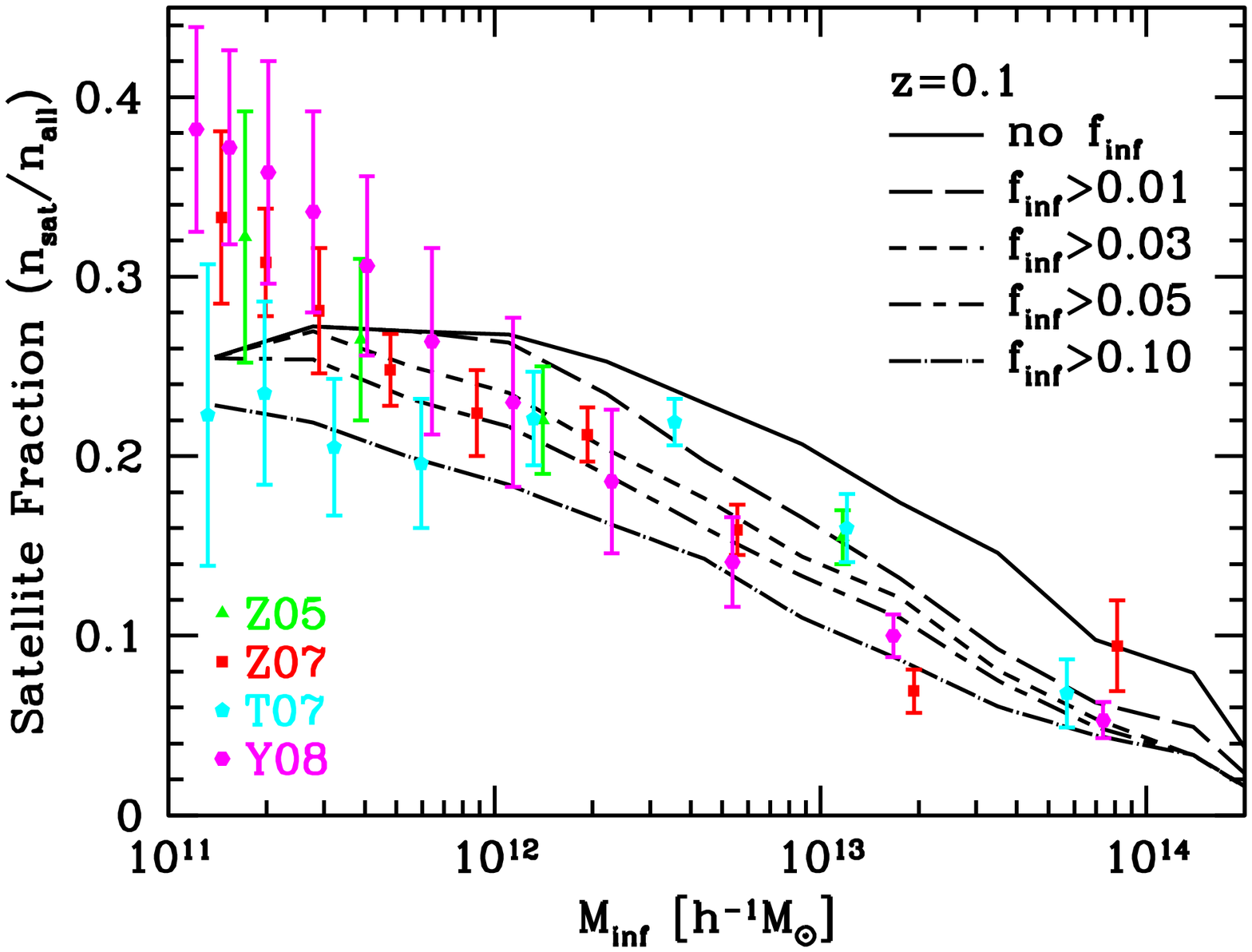}}
\resizebox{3.3in}{!}{\includegraphics{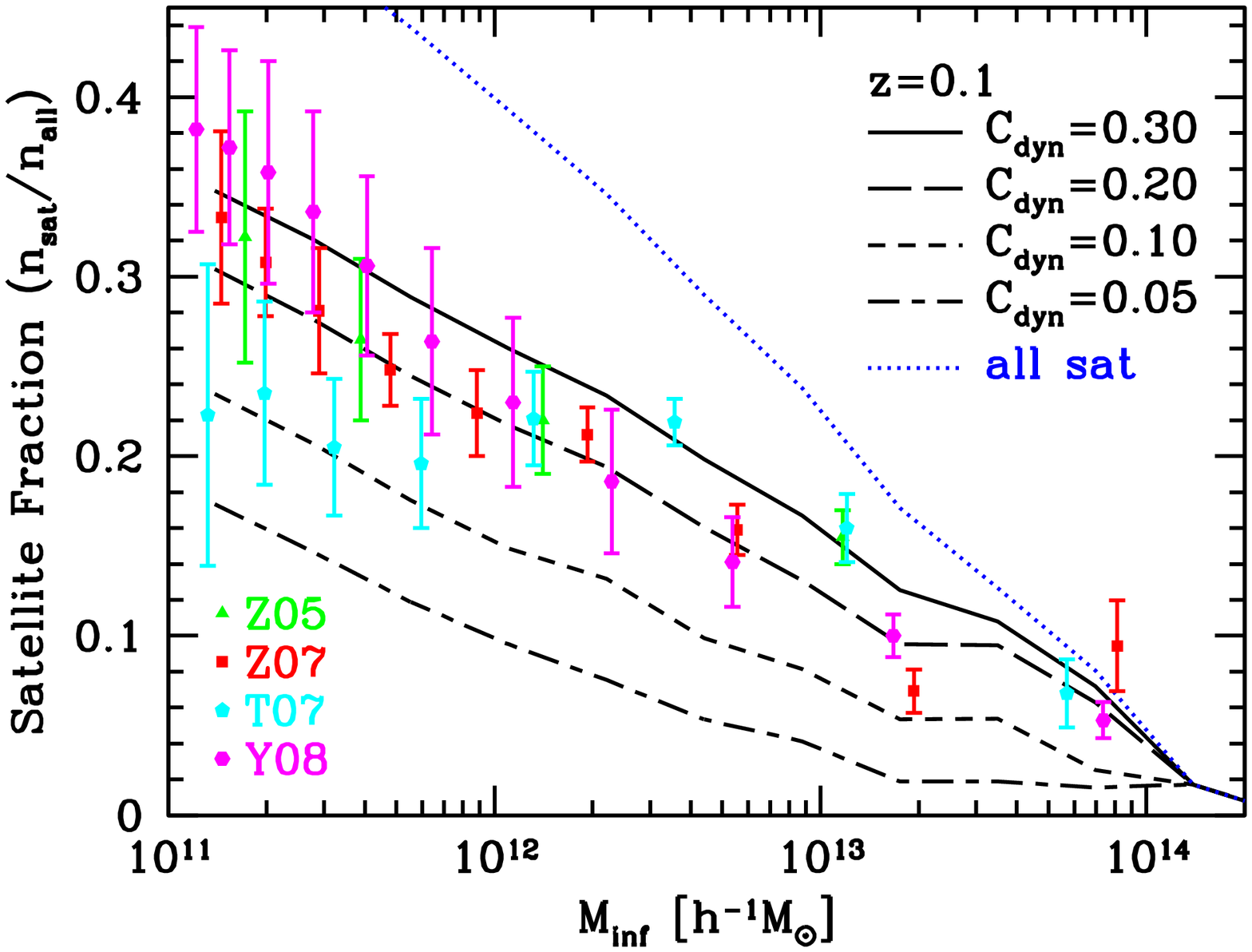}}
\end{center}
\vspace{-0.1in}
\caption{
Satellite fraction vs. infall mass.
\textbf{Top}: Using various removal thresholds, $f_{\rm inf}$, from the subhalo catalog.
Points show observationally derived values, matched in number density to our subhalo catalog, from \citet{ZehZheWei05} (Z05), \citet{ZheCoiZeh07} (Z07), 
\citet{TinNorWei07} (T07), and \citet{YanMovdB08a} (Y08).
\textbf{Bottom}: Using the analytic model for satellite halo infall of Eq.~\ref{eq:dynfric} and various values of $C_{\rm dyn}$ for dynamical friction.
Good agreement with observation (no rollover) continues at low mass since the analytic model depends only on resolving halos prior to infall.
Dotted curve shows keeping all satellites ever accreted.
} \label{fig:satfrac}
\end{figure}

Figure~\ref{fig:satfrac} (top) shows observed satellite fractions, abundance matched to our subhalo catalog and plotted as a function of $M_{\rm inf}$.
The observationally-based analyses yield similar results, which also agree with values from galaxy-galaxy lensing \citep{ManSelKau06}, though there is considerable scatter.
At the low mass end, the amplitude from \citet{TinNorWei07} is anomalously low, which may be driven by the fact that the 2dFGRS is a blue-selected samples and thus could be missing a large population of satellites there, since satellites are preferentially redder than centrals \citep{vdBAquYan08}.
Given the consistency of the three other samples at low mass, we consider their values more robust in that regime.
At the high mass end, the satellite fractions from \citet{ZheCoiZeh07} show an upturn; since the satellite fraction is expected to decrease with mass and the amplitude of the highest two mass bins are considerably different than the other samples, we will not consider them as robust data points.

Also shown are satellite fractions from our subhalo catalog for various removal thresholds, $f_{\rm inf}$.
Since the satellite fraction should rise with decreasing galaxy mass, the rollover at low mass clearly shows the limits of numerical resolution below $M_{\rm inf}=10^{11.5}\,h^{-1}M_\odot$.
Above this mass, $f_{\rm inf} \approx 0.03$ works well across over two decades in $M_{\rm inf}$, consistent with results from spatial clustering.

Note that the satellite fraction remains non-zero up to $M_{\rm inf} \sim 10^{14}\,h^{-1}M_\odot$, resulting from low redshift cluster-cluster mergers \citep[e.g.][]{CohWhi05}.
This implies that $\sim10\%$ of galaxy clusters should be expected to host a satellite BCG, as supported by observations which shown a number of clusters hosting BCG-BCG pairs \citep[e.g.,][]{LiuMaoDen09}.
In a sample of rich clusters from the SDSS MaxBCG catalog, $15\%$ host two BCGs \citep{KoeMcKAnn07a}.

We can also use our analytic model for satellite removal as in \S\ref{sec:dynfric} to examine how different $C_{\rm dyn}$ values for dynamical friction affect the satellite fraction, as shown in Fig.~\ref{fig:satfrac} (bottom).
Discounting the results of \citet{TinNorWei07} at low mass, which exhibit a flattening that is not seen in any of the other samples, our simple analytic model provides good agreement with observations using $C_{\rm dyn}=0.2-0.3$.
This provides a consistent picture, since using $C_{\rm dyn}=0.2-0.3$ matches the subhalo HOD using $f_{\rm inf}=0.01-0.03$ (Fig.~\ref{fig:hodtdyn}), which is also the range of $f_{\rm inf}$ that best matches observed clustering.
Additionally, this analytic model works well even below our mass threshold for reliably tracking satellite subhalos, showing no rollover at low mass since it depends only on resolving halos prior to infall.

The dotted curve in Fig.~\ref{fig:satfrac} (bottom) shows the satellite fraction if satellites are never removed (merge with central or tidally disrupt), which we measure by analytically retaining every infalling halo, as in Fig.~\ref{fig:hod}.
The slight deficit compared with the full subhalo catalog at high mass arises because satellites in this analytic model do not gain mass via merging with other satellites.
While this ``no removal'' scenario provides a reasonable estimate at high mass, where halos have formed only recently and so their average times since infall are short, it grossly overpredicts the satellite fraction at lower mass.
This demonstrates the clear failure of a scenario in which satellite galaxies never merge/disrupt.

Overall, we find that $C_{\rm dyn} \approx 0.25$ both matches well to the satellite removal criteria that agrees with spatial clustering, and it directly produces satellite fractions in agreement with observations.
\citet{ConHoWhi07} apply a similar dynamical infall model as ours to halo merger trees to examine the growth of the Luminous Red Galaxy (LRG) population.
They find that $C_{\rm dyn}=0.1$ fit observations well, though they were unable to put precise constraints on this value.
Figure~\ref{fig:satfrac} (bottom) shows that such a value would underpredict the observed satellite fraction across a broad mass range.

$C_{\rm dyn}=0.2$ also agrees well with a similar analysis by \citep{WetCohWhi09a}, who match the evolution of satellite subhalos in simulations at $z \gtrsim 1$.
They also pointed out that, given typical satellite orbital circularity distributions \citep{ZenBerBul05,JiaJinFal08}, the detailed fits to dynamical friction infall times of \citet{BoyMaQua08} and \citet{JiaJinFal08} predict $C_{\rm dyn}=0.06$ and $0.14$, respectively.
The latter value, obtained from an analysis of a cosmological simulation not dissimilar from ours, is marginally consistent with our results.
However, the former clearly is not.
This can partially be attributed to the different mass dependence in the parametrization of \citet{BoyMaQua08}: they used a fit similar to Eq.~\ref{eq:dynfric} but with $(M_{\rm halo}/M_{\rm sat,inf})^{1.3}$ in the numerator.
Their added exponent factor doubles their merging/disruption time (with respect to ours) for a $10:1$ mass ratio merger.
While this does provide better agreement, it still leads to a $10:1$ infall timescale that is half of ours.
Given that their fit was obtained from an analysis of significantly higher resolution mergers of two isolated, spherical (Hernquist profile) halos, this highlights the importance of realistic cosmological settings for calibrating satellite merging/disruption timescales.

Finally, we note that the above analytic model of satellite removal is dependent on halo definition.
In particular, if halos are defined to have a larger radius, then satellites will accrete sooner and hence the analytic removal time will be longer.
As compared with the commonly used FoF($b=0.2$), $r_{b=0.168} \approx 0.86\,r_{b=0.2}$.
Thus, the best-match $C_{\rm dyn} \approx 0.25$ values represents a lower limit to that of satellites in halos using FoF($b=0.2$).\footnote{
In principle, changing halo definition can also change the mass ratio in Eq.~\ref{eq:dynfric} and hence the removal time, since halo concentration scales weakly with mass.
However, the mass enhancement from FoF($b=0.168$) to FoF($b=0.2$) is nearly independent of concentration and redshift.
The relation to spherical overdensity halos will depend on concentration and redshift \citep[see][for more on halo conversion]{Whi01}.}

\subsection{Cluster Satellite Luminosity Function} \label{sec:clf}

Since the correlation function and satellite fraction conflate satellites within a variety of host halo masses, we finally test our subhalo catalog by examining satellites specifically in the densest environments: galaxy clusters.
We compare the luminosity function of satellites within galaxy clusters in our simulation with that determined from the SDSS MaxBCG catalog \citep{HanSheWec09}.
Here, we abundance match our subhalo catalog at $z=0.25$ to the SDSS $i$-band galaxy luminosity function at that redshift \citep{SheJohMas09}, from which the cluster catalog in \citet{HanSheWec09} is drawn.
Additionally, we assume $0.15$ dex scatter in luminosity at fixed subhalo mass.
We then compute the luminosity function of satellites in clusters of a given richness, where we define richness as in \citet{HanSheWec09} as the number of red sequence satellite galaxies with $L>0.4L_*$ within the cluster virial radius.
We use their fit to the red galaxy fraction vs. cluster richness (their Eq.~13) to scale our total satellite counts to those of red galaxies (this fraction is $65-80\%$ in the richness range we probe).

\begin{figure}
\begin{center}
\resizebox{3.3in}{!}{\includegraphics{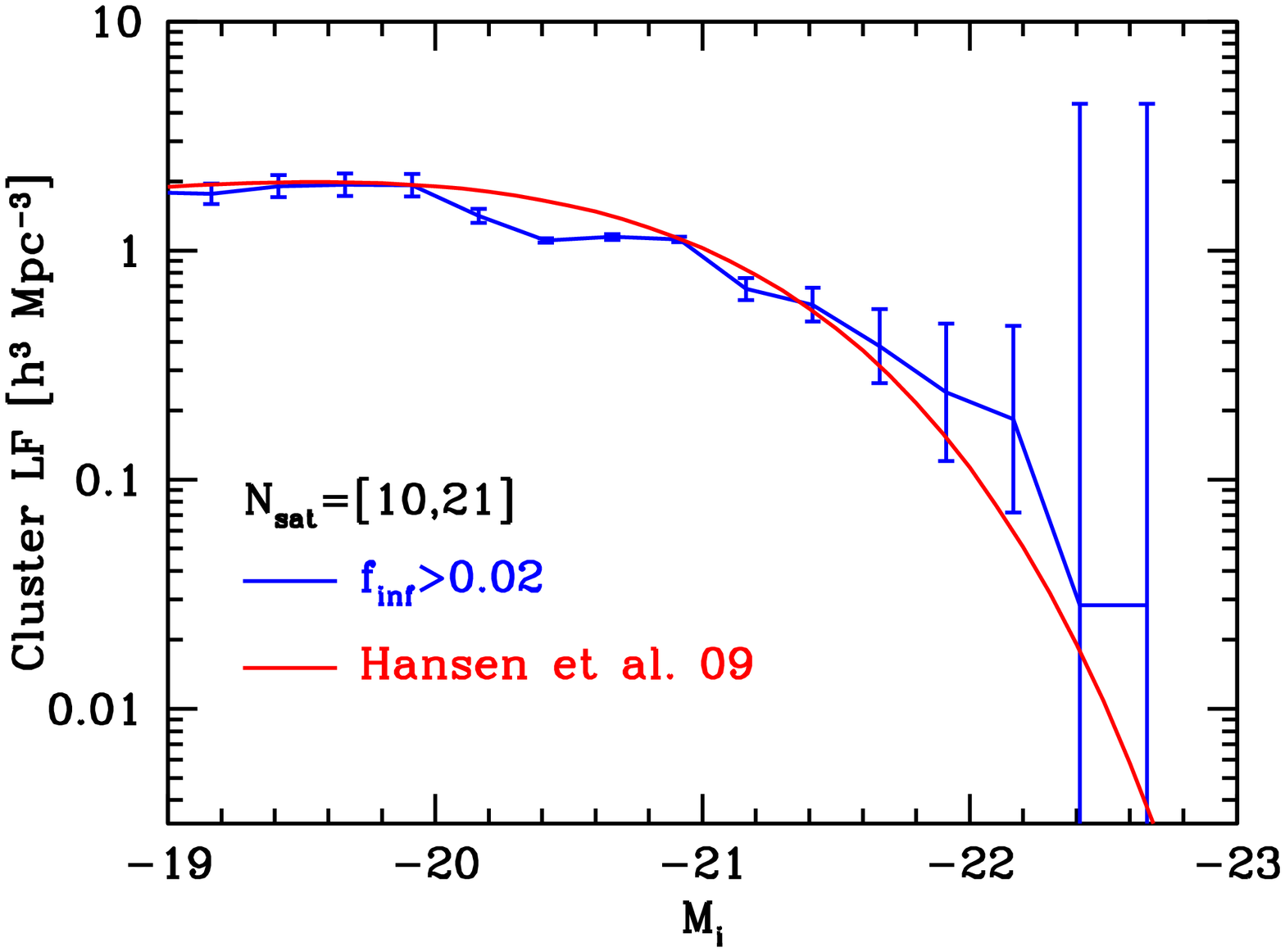}}
\resizebox{3.3in}{!}{\includegraphics{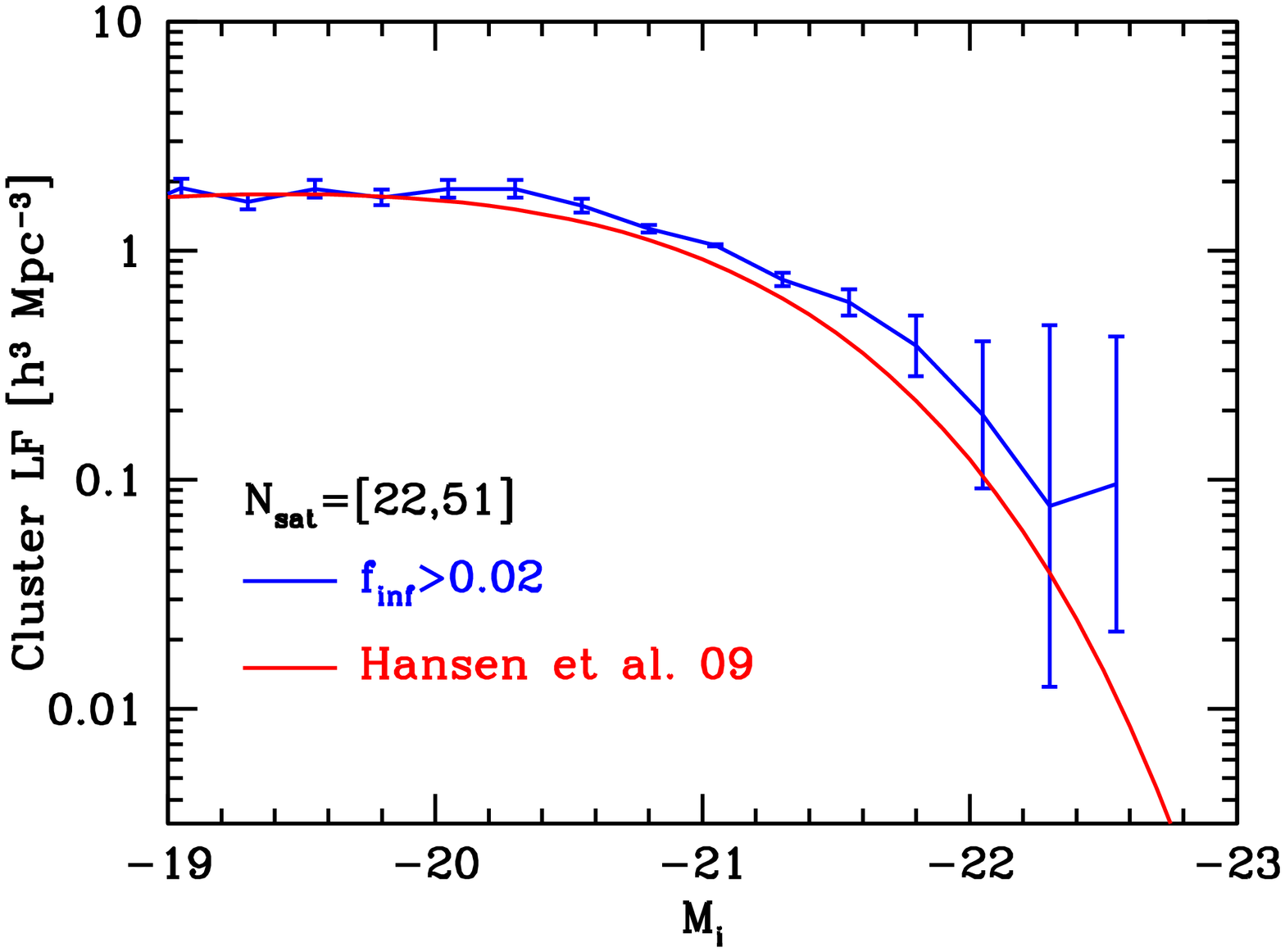}}
\end{center}
\vspace{-0.1in}
\caption{
Luminosity function of satellites in galaxy clusters at $z=0.25$, for lower (top) and higher (bottom) cluster richness bins.
Subhalos are abundance matched to the SDSS $i$-band luminosity function \citep{SheJohMas09}, assuming $0.15$ dex scatter in luminosity at fixed subhalo mass, and satellites retained if their $f_{\rm inf}>0.02$.
Solid curve shows fit to SDSS MaxBCG catalog from \citet{HanSheWec09}.
} \label{fig:clf}
\end{figure}

Figure~\ref{fig:clf} shows the results of our subhalo catalog, using a removal threshold of $f_{\rm inf}>0.02$, as compared with \citet{HanSheWec09}.
We find good agreement for both high and low bins of cluster richness.
Since $M_i=-19$ corresponds to a subhalo infall mass of $10^{11.4}\,h^{-1}M_\odot$, the agreement in Fig.~\ref{fig:clf} across a wide range of luminosities demonstrates that our subhalo catalog, with a model for merging/disruption, successfully traces the galaxy population in the densest environments down to our expected numerical resolution threshold.
Figure~\ref{fig:clf} does exhibit some excess of very bright cluster satellites in our catalog, though the limited number of massive clusters in our simulation volume precludes a definitive test in that regime.

\section{High Redshift} \label{sec:highz}

\subsection{HOD and Analytic Removal} \label{sec:hodhighz}

Examining the effects of removal thresholds on the HOD at higher redshift, we find little change in the fractional reduction in the satellite HOD as $f_{\rm inf}$ is increased.
This supports a picture where satellite subhalo mass stripping is quite rapid after infall, such that the shorter average times since infall at higher redshift do not lead to a significantly less stripped satellite subhalo population.

\begin{figure}
\begin{center}
\resizebox{3.3in}{!}{\includegraphics{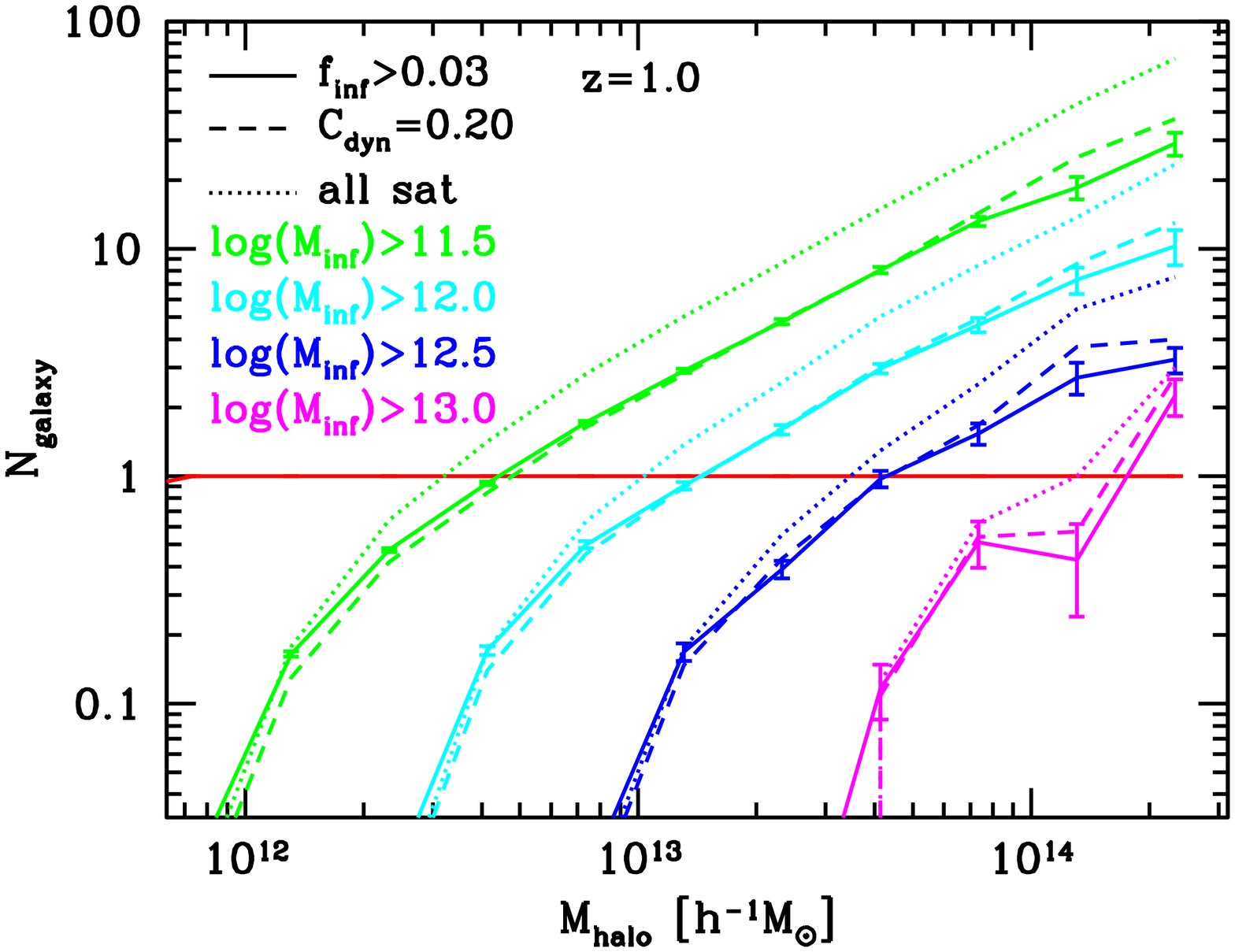}}
\resizebox{3.3in}{!}{\includegraphics{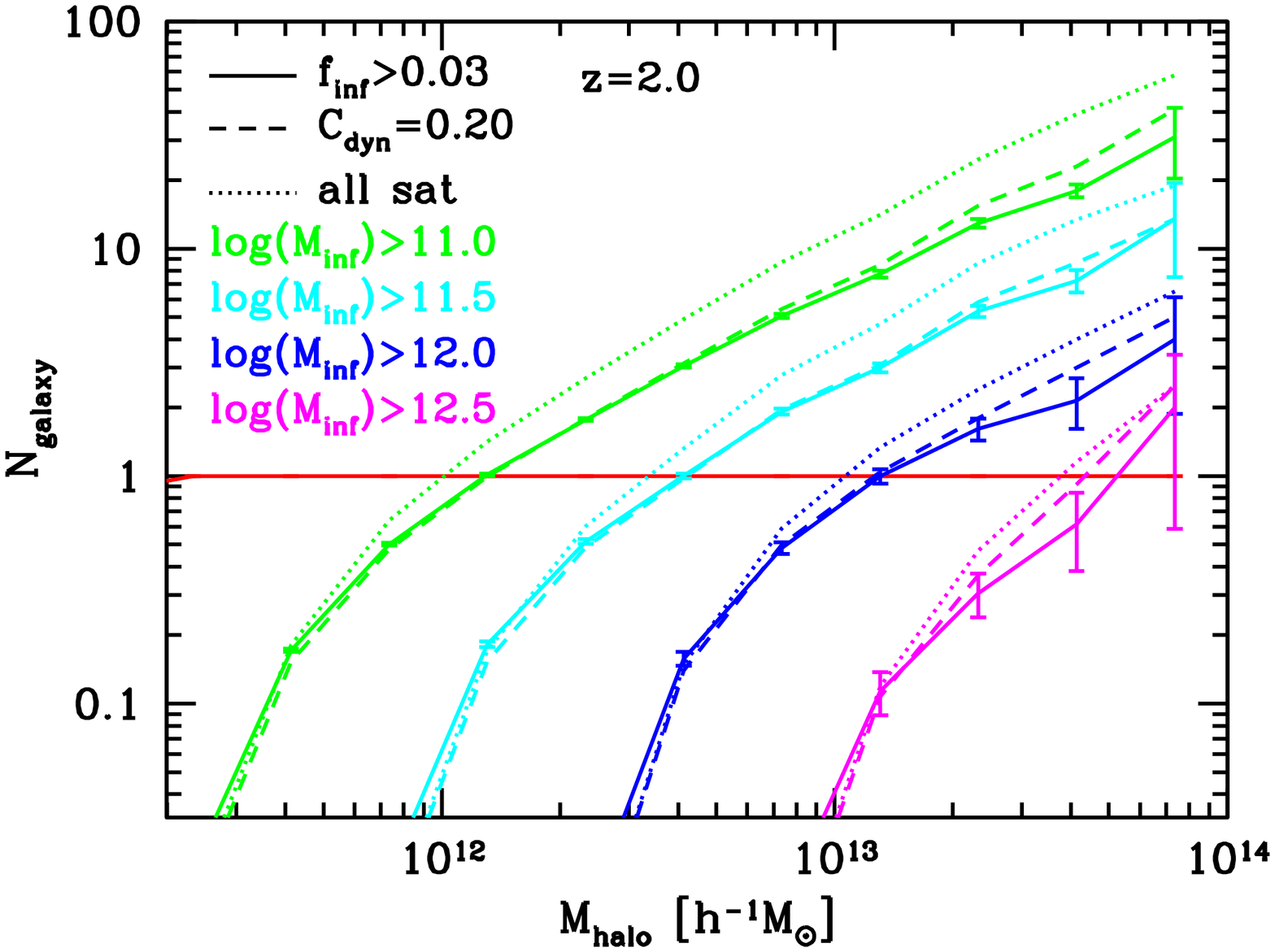}}
\end{center}
\vspace{-0.1in}
\caption{
Same as Fig.~\ref{fig:hodtdyn}, but at $z=1$ (top) and $z=2$ (bottom).
A given $C_{\rm dyn}$ value maps nearly as well to a given removal threshold, $f_{\rm inf}$, at high $z$ as at $z=0.1$.
Dotted curves show keeping all satellites ever accreted.
At higher $z$, this case exhibits less excess with respect to the dynamical infall prescription.
} \label{fig:hodtdynhighz}
\end{figure}

We also examine our analytic model for satellite removal timescale (Eq.~\ref{eq:dynfric}) at higher redshift.
In particular, we test the validity of the scaling of $t_{\rm dyn}$ with $t_{\rm Hubble}$.
Figure~\ref{fig:hodtdynhighz} shows the HOD at $z=1$ and $2$ for both the subhalo catalog and analytic model, using the same $f_{\rm inf}$ and $C_{\rm dyn}$ values as in Fig.~\ref{fig:hodtdyn} (bottom).
The agreement at all host halo and satellite masses remains relatively robust given the simplicity of the model.
This agreement is partially a result of the decreasing dependence of the analytical model on $C_{\rm dyn}$ at higher $z$, when halos of a given mass have formed and been accreted more recently, so allowing longer infall times does not significantly increase the number of satellites.
This is demonstrated by decreasing difference in the HOD for $C_{\rm dyn}=0.2$ (dashed curves) as opposed to the ``no removal'' scenario (dotted curves).
Moreover, using $C_{\rm dyn} \gtrsim 0.3$ produces no additional enhancement in the HOD at $z>2$.

This result has clear implications for modelling galaxy evolution at high redshift.
Since the details in modeling satellite galaxy infall matter less at higher redshift, nearly all uncertainty in understanding galaxy evolution at high redshift lies in the gas/radiation physics and not in the dynamics of dark matter.

\subsection{Spatial Clustering at $z\sim1$} \label{sec:wphighz}

\begin{figure}
\begin{center}
\resizebox{3.3in}{!}{\includegraphics{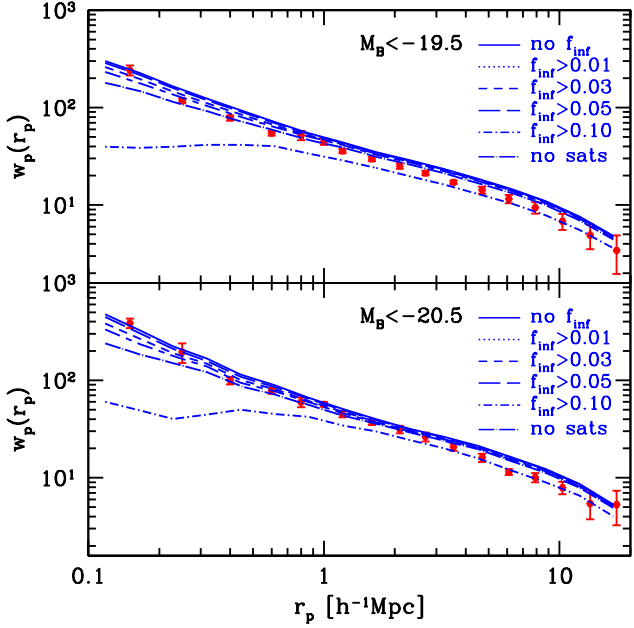}}
\end{center}
\vspace{-0.1in}
\caption{
Projected auto-correlation function for several removal thresholds, $f_{\rm inf}$, as compared with the observed clustering at $z\sim1$ of \citet{CoiNewCoo06}.
$M_B<-19.5$, $-20.5$ samples correspond to subhalo $M_{\rm inf}\gtrsim10^{11.5}$, $10^{12.0}\,h^{-1}M_\odot$.
Abundance matching here assumes $0.6$ dex scatter in $L_B$ at fixed $M_{\rm inf}$.
} \label{fig:wphighz}
\end{figure}

We also examine whether our criteria for satellite removal hold at higher redshift by comparing with galaxy clustering results at $z\sim1$ from the Deep Extragalactic Evolutionary Probe 2 (DEEP2) survey \citep{CoiNewCoo06}, as shown in Fig.~\ref{fig:wphighz}.
Here, we use abundance matching to match the number densities of $B$-band luminosity threshold galaxy samples in DEEP2.
We find that using abundance matching with no scatter in the $L_B-M_{\rm inf}$ relation causes our subhalo catalog to exhibit significant excess in clustering as compared with DEEP2 for all but the lowest few $r_p$ bins.
We find improved agreement using significant ($0.6$ dex) scatter, as shown for two luminosity thresholds corresponding to subhalos above our robust resolution limit.
Using even larger scatter improves the agreement on large scales, but it underpredicts clustering on smaller scales.

At $z\sim1$, the small scale ($r_p<1\,h^{-1}$~Mpc) clustering provides a consistent picture with the results at $z=0.1$, favoring a removal threshold of $f_{\rm inf}=0.01-0.05$.
However, despite the large luminosity-mass scatter, our subhalo catalog overpredicts the clustering as compared with DEEP2 on large scales.
Curiously, at face value, DEEP2 clustering at $r_p \sim 10\,h^{-1}$~Mpc favors a model with \textit{no\/} satellite galaxies.

There are a number of possible reasons for why DEEP2 suggests increased scatter and/or reduced clustering.
DEEP2 selects on $B$-band luminosity, which is more susceptible to recent star formation, and since satellites at $z\sim1$ are redder than centrals at a given stellar mass \citep{ZheCoiZeh07,TinWet09}, this could bias the clustering low since satellites preferentially live in more massive halos, which are more highly biased.
As discussed in \S\ref{sec:minf}, it is possible that the $L-M_{\rm inf}$ relation has more scatter at higher redshift.
However, large scatter alone is unable to match the clustering at all scales.
Clustering in DEEP2 could be biased low because their selection criteria miss $\sim10\%$ of red galaxies in their survey volume, although this effect is likely to be mild both because red galaxies form the minority of the population at all luminosity thresholds and because $w_p(r_p)$ is nearly the same for red and blue galaxies at $r_p \sim 10\,h^{-1}$~Mpc \citep{CoiNewCro08}.
Alternately, if DEEP2 is missing an appreciable number of galaxies in their survey volume (regardless of color), our abundance matching method will correlate to the observed galaxies artificially high subhalo masses, leading to enhanced clustering.
Or the disagreement could be a statistical fluctuation.
The DEEP2 survey volume of $\sim$($100\,h^{-1}$~Mpc)$^3$ is susceptible
to sampling variance, and the clustering at different scales are highly correlated, so the discrepancy may not be statistically significant.
Since we do not have a covariance matrix for the observations, we are unable
to determine the goodness-of-fit.

Interestingly, \citet{ConWecKra06} find good agreement with DEEP2 clustering using similar method based on abundance matching against subhalo catalogs.
Their better agreement (lower clustering) may stem from their different assumed cosmology ($\sigma_8=0.9$), and/or their simulation box sizes ($80$ and $120\,h^{-1}$~Mpc), significantly smaller than the one used here.
Suggestively, a recent HOD analysis using color information finds a similar clustering trend as ours, namely, good agreement at small scales but excess clustering at $r_p\gtrsim1\,h^{-1}$~Mpc \citep{TinWet09}.

Thus, while our satellite removal criteria are consistent with clustering at $z\sim1$ in DEEP2 at $r_p<1\,h^{-1}$~Mpc given $L_B-M_{\rm inf}$ scatter, the above uncertainties and lack of agreement in large-scale clustering do not allow us to obtain strong constraints from these higher redshift data.

\subsection{Evolution of the Satellite Fraction} \label{sec:satfrachighz}

\begin{figure}
\begin{center}
\resizebox{3.3in}{!}{\includegraphics{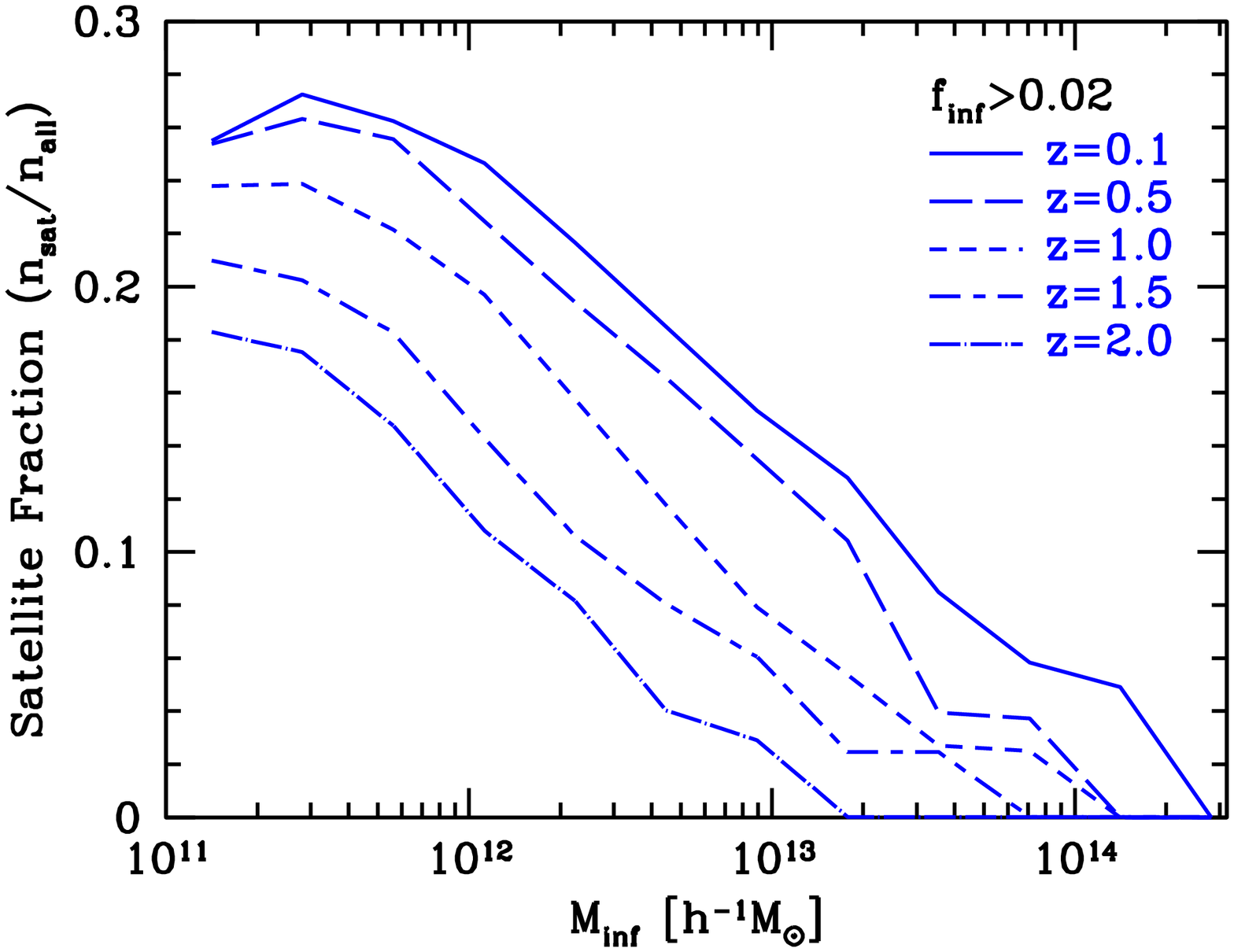}}
\resizebox{3.3in}{!}{\includegraphics{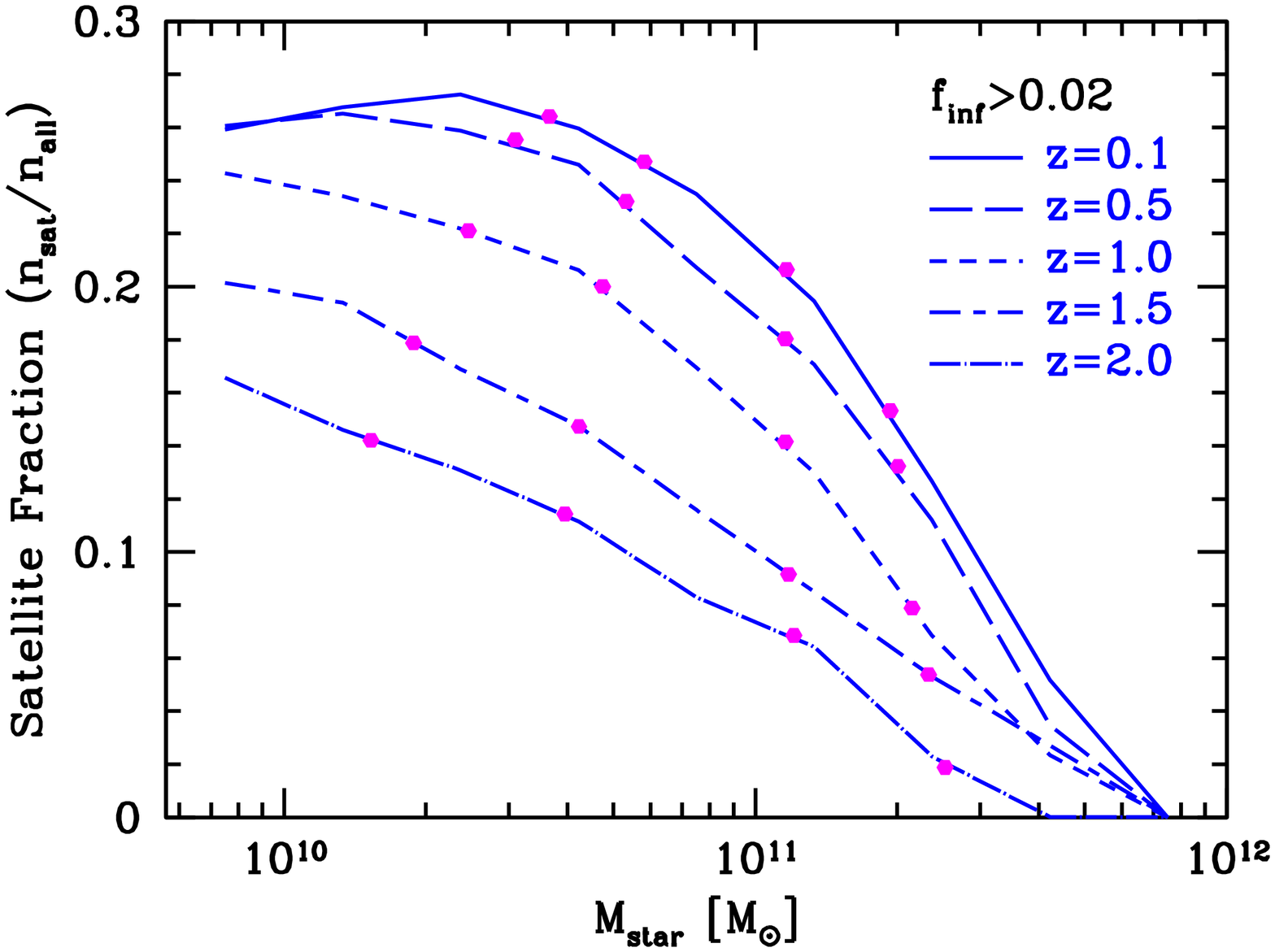}}
\end{center}
\vspace{-0.1in}
\caption{
Satellite fraction vs. mass at various redshifts, using a removal threshold of $f_{\rm inf}=0.02$.
\textbf{Top}: Subhalos binned in $M_{\rm inf}$.
The turnover in satellite fraction occurs at lower $M_{\rm inf}$ at higher $z$, indicating that resolution effects less critical at higher $z$.
\textbf{Bottom}: Subhalos binned in $M_{\rm star}$, based on abundance matching.
Points indicate values at fixed $M_{\rm inf}$ of $10^{11.75},10^{12.0},10^{12.5},10^{13.0}\,h^{-1}M_\odot$, highlight the evolving $M_{\rm star}-M_{\rm inf}$ relation.
} \label{fig:satfracevol}
\end{figure}

Finally, we explore how the satellite fraction evolves with redshift.
Figure~\ref{fig:satfracevol} (top) shows the satellite fraction as a function of $M_{\rm inf}$ at various redshifts, using $f_{\rm inf}=0.02$.
At a fixed infall mass, the satellite fraction monotonically increases with time, arising from the decreasing dynamical friction infall time with respect to the typical halo merger timescale \citep{WetCohWhi09a}.
The evolution of the rollover at low mass, arising from numerical disruption, indicates that the resolution limit decreases with higher redshift, when satellites have been accreted more recently.

Since the $M_{\rm star}-M_{\rm inf}$ relation evolves with time, Fig.~\ref{fig:satfracevol} (bottom) shows the satellite fraction instead as a function of $M_{\rm star}$, obtained by abundance matching against the SMF of \citet{ColNorBau01} at $z=0.1$ and \citet{MarvDoFor09} at higher redshift.
Points show the satellite fraction at fixed $M_{\rm inf}$ values.
At high mass, the $M_{\rm star}-M_{\rm inf}$ relation exhibits only mild evolution, but at lower mass, subhalos at a fixed $M_{\rm inf}$ grow rapidly in $M_{\rm star}$.

This result has two important implications.
First, in examining a galaxy population of fixed stellar mass, simulation resolution becomes even less of a limitation at higher redshift than Fig.~\ref{fig:satfracevol} (top) would indicate, since at the low mass end galaxies of a fixed $M_{\rm star}$ move to a higher $M_{\rm star}$ with redshift.
This is evidenced by the lack of rollover in the stellar mass satellite fraction at higher redshift.

Second, the growth of $M_{\rm star}$ at fixed $M_{\rm inf}$ for $M_{\rm star}<10^{11}M_\odot$ demonstrates how SHAM implicitly allows for satellite star formation.
For example, if a satellite were accreted at $z=1$ with $M_{\rm star}=4\times10^{10}M_\odot$, following the points of fixed $M_{\rm inf}$ in Fig.~\ref{fig:satfracevol} (bottom) shows that SHAM-assigned stellar mass would increase by $\sim20\%$ by $z=0.1$.

\section{Comparisons with Other Removal Criteria} \label{sec:comparison}

As discussed in the introduction, various authors have used different criteria to define satellite galaxy removal via dark matter subhalos.
Here, we examine how other criteria compare against our fiducial case, the subhalo bound mass to infall mass ratio, $f_{\rm inf}$.

As shown by our full tracking scheme (keeping subhalos to $50$ particles), using a fixed mass criterion can reasonably match observations if the mass threshold corresponds to $1-5\%$ of the infall mass.
As simulations increase in resolution, it thus becomes increasingly necessary to model for satellite removal before subhalos reach numerical disruption.

Models which remove a satellite subhalo after its mass has been stripped to a value less than the mass within a fixed fraction ($a_{\rm dis}$) of its NFW scale radius \citep[e.g.,][]{TayBab04,ZenBerBul05,MacKanFon09} are largely degenerate with our fiducial $f_{\rm inf}$ criterion, since subhalo concentration scales only weakly with mass.
For a typical halo concentration of $c=10$, our results suggest $a_{\rm dis} \approx 0.3$, within the range $0.1-1$ often assumed.

Given that total baryonic to dark mass ratio in a halo is nearly $10\%$, 
\citet{SteBulBar09} assumed that a satellite merges with its central galaxy when it drops below $10\%$ of infall mass.
However, we find that $f_{\rm inf}=0.1$ noticeably underpredicts the satellite population.
A lower threshold is needed, likely both because the stellar to dark mass ratio is significantly lower ($\sim4\%$), and because any baryonic mass will be more compact than the $10\%$ most bound dark mass.

Since the $M_{\rm star}/M_{\rm inf}$ ratio varies with subhalo mass (Fig.~\ref{fig:mstel}), it is not obvious that our best-fit removal threshold, based on subhalos of $M_{\rm inf}>10^{11.5}\,h^{-1}M_\odot$ ($L>0.2\,L_*$), can be simply extrapolated to modelling dwarf-spheroidal systems around the Milky Way.
However, our results are consistent with allowed dark mass stripping fractions to match the Milky Way population \citep{PenNavMcC08,MacKanFon09}.

Instead of mass stripping-based criteria, we also examine criteria based on satellite tidal radius or distance from halo center/central galaxy.
To obtain precise central-satellite separations, we linearly extrapolate satellite positions between subsequent outputs to find the minimum separation.
For succinct comparison, we compare satellite fractions against $f_{\rm inf}=0.02$, the fiducial criterion which matches well to observations as shown above.

\begin{figure}
\begin{center}
\resizebox{3.3in}{!}{\includegraphics{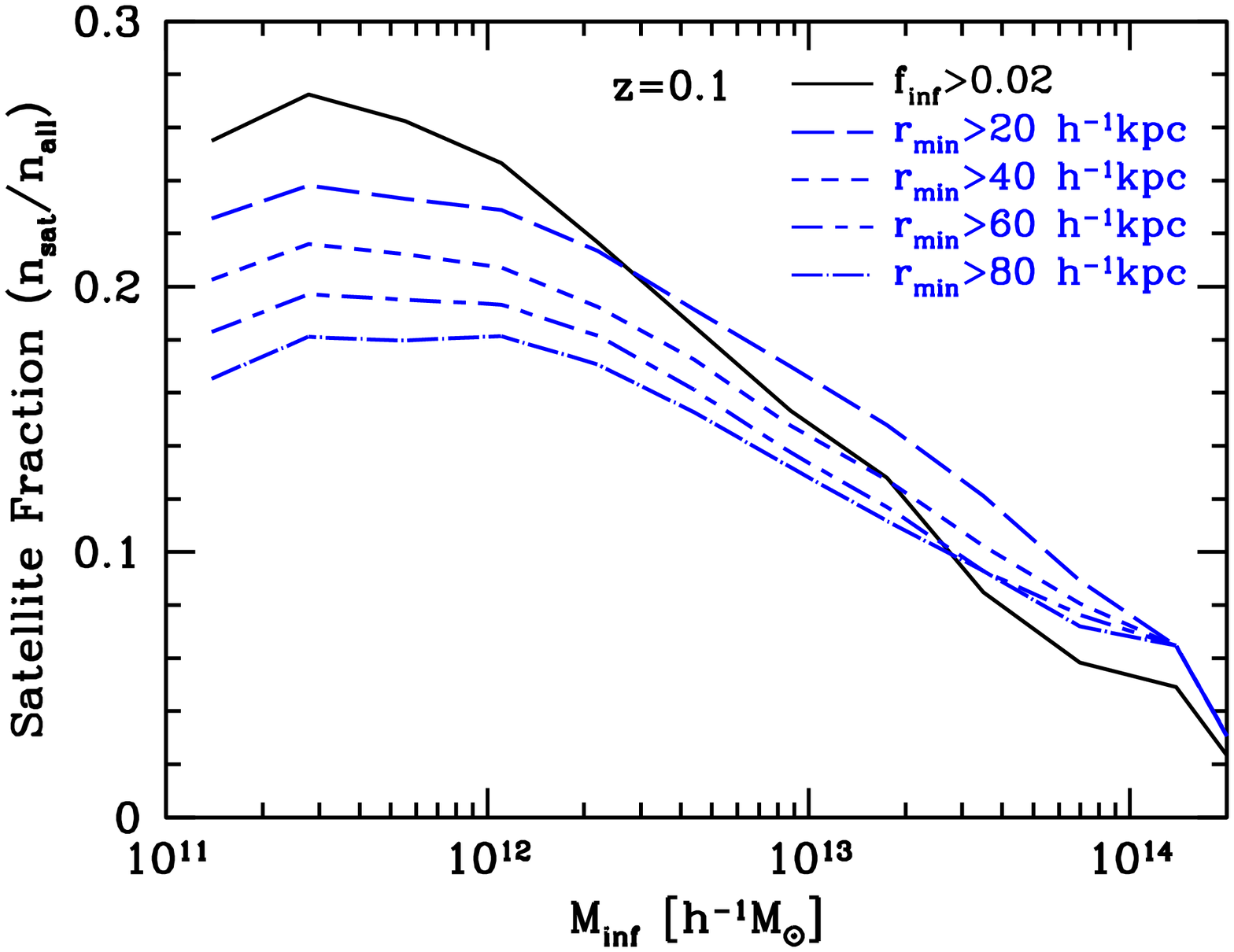}}
\resizebox{3.3in}{!}{\includegraphics{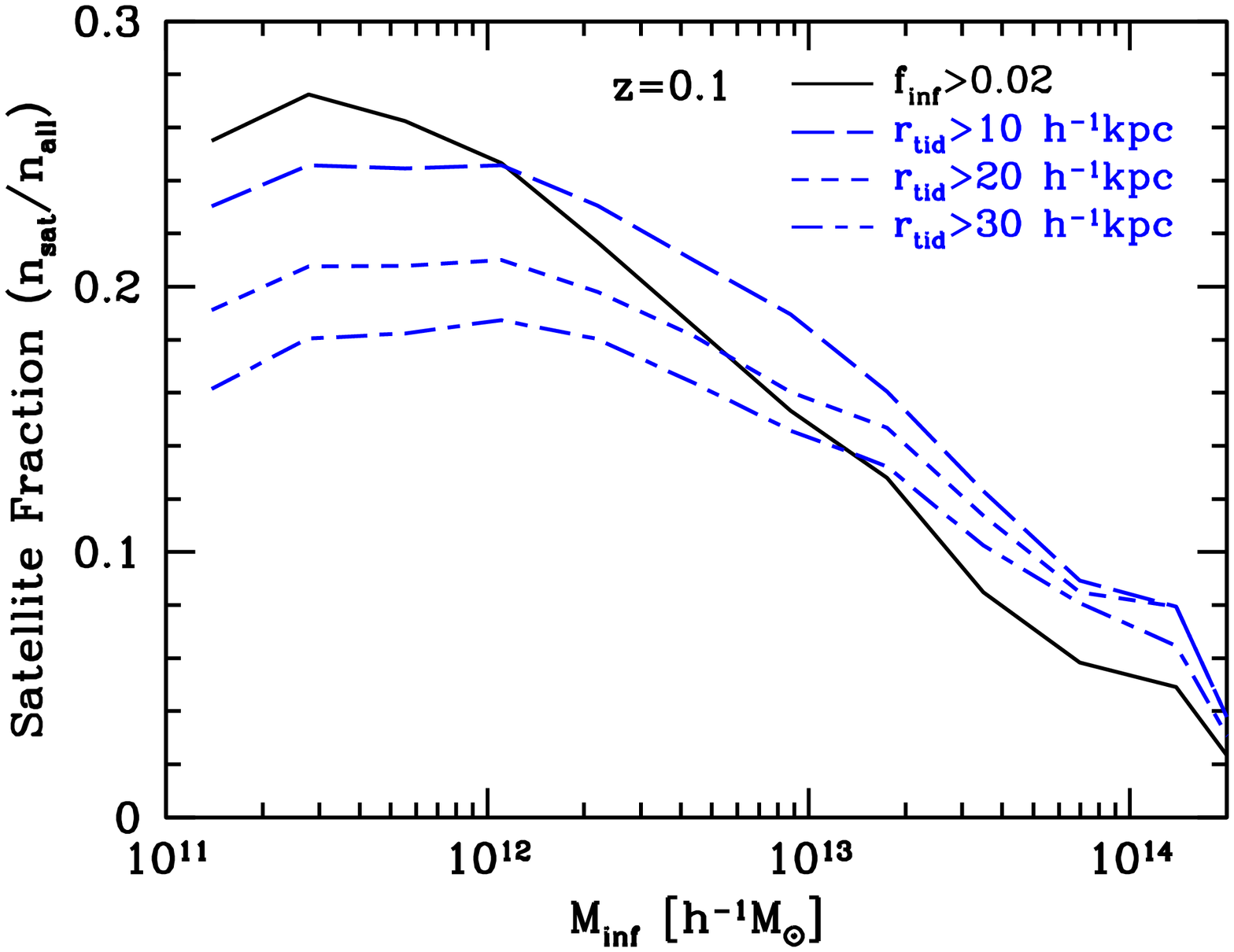}}
\resizebox{3.3in}{!}{\includegraphics{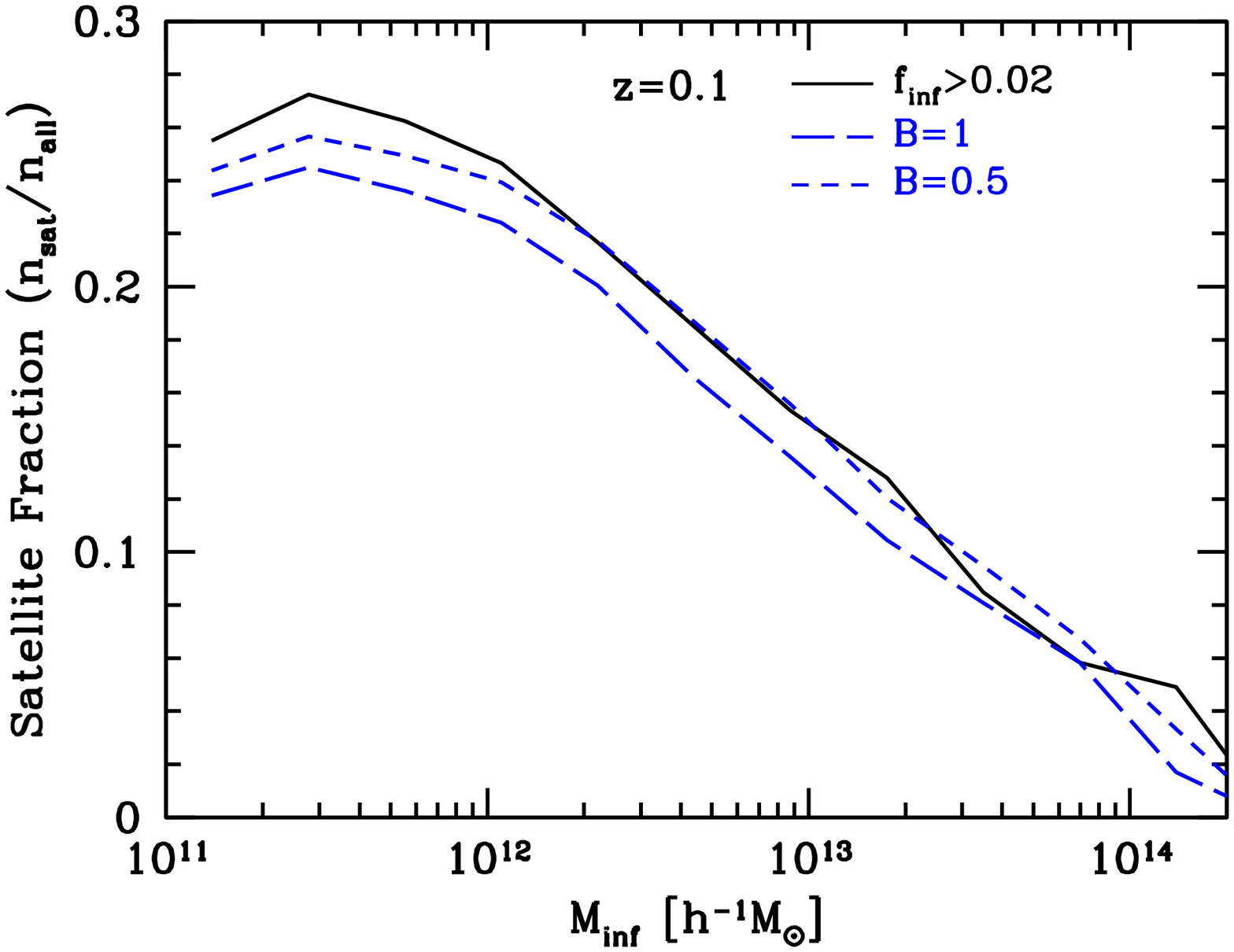}}
\end{center}
\vspace{-0.1in}
\caption{
Satellite fraction vs. subhalo infall mass, using various removal criteria.
\textbf{Top}: Satellite removed when it passes within a given physical distance of its central galaxy.
\textbf{Middle}: Satellite removed when its tidal radius falls below the given value.
\textbf{Bottom}: Satellite removed when it passes within radius at which both satellite subhalo and host halo have same interior mass (ignores satellite mass stripping), both using NFW profiles from simulation and approximating halos as isothermal spheres.
} \label{fig:satfracother}
\end{figure}

As was shown in Fig.~\ref{fig:dist}, a correlation exists between a satellite's distance from its central galaxy and falling below a given $f_{\rm inf}$ value.
Thus, Fig.~\ref{fig:satfracother} (top) shows satellites removed when coming within a fixed physical distance from the central galaxy.
Despite the stripping-distance correlation, compared with our fiducial case this model underpredicts satellites at low mass and overpredicts at high mass.
Combined with the results from our full tracking model (fixed mass threshold), this implies that any successful removal criteria must scale with the mass or radius of the satellite subhalo.

In this vein, the middle panel shows satellite removal if a satellite subhalo's tidal (Jacobi) radius falls below a given value, where
\begin{equation} \label{eq:tidal}
  r_{\rm tid} = A\left[\frac{M_{\rm sat}}{3M_{\rm halo}(<r)}\right]^{1/3}\,r
\end{equation}
and where $r$ is the physical separation from satellite to halo center, $M_{\rm sat}$ is the instantaneous bound satellite mass, $M_{\rm halo}(<r)$ is the halo mass within $r$, and $A$ is a constant that accounts for the host halo's density profile \citep[see][Eq.~8.107, 8.108]{BinTre}.
Assuming that a satellite disrupts when within its host halo's inner NFW profile lets us set $A=1.4$, and we measure $M_{\rm sat}$ and $M_{\rm halo}(<r)$ directly from our simulation.
As a simple analytic alternative, we have investigated modelling the halos at infall as isothermal spheres \citep[as applied to][Eq.~9.86]{BinTre}, which produces similar results.
Figure~\ref{fig:satfracother} (middle) shows the satellite fraction for tidal radii motivated by typical galaxy sizes.
Similar to using fixed satellite distance, this method predicts too shallow of a dependence on satellite infall mass, likely a manifestation of its weak ($1/3$ power) scaling with mass ratio.

Finally, we consider a removal criterion defined simply by a linear scaling relation between satellite halo -- host halo mass and radius at infall.
Specifically, removal occurs when
\begin{equation}
  r < B\left(\frac{M_{\rm sat,inf}}{M_{\rm halo}}\right)R_{\rm halo,vir}
\end{equation}
where $B$ is some constant.
Note that if satellite and host halos are isothermal spheres, such that $M(<r)/M_{\rm vir}=r/R_{\rm vir}$, and $B=1$, the above criterion is met when the halo mass interior to $r$ equals the satellite's infall mass.
Figure~\ref{fig:satfracother} (bottom) shows the results using $B=1$ and $0.5$.
This model provides good agreement with our fiducial case because of the strong (linear) scaling with mass ratio, with $B=0.5$ well matching the overall amplitude.
Thus, this method may provide a useful approximation for simple analytic models of satellite evolution.

Various works using semi-analytic models grafted onto subhalo populations use an analytic model to track satellite galaxies after subhalo disruption in order to match small-scale galaxy clustering at $z\sim0$ \citep[e.g.,][]{SprWhiTor01,KitWhi08,SarDeLDol08}.
While this method does provide a means to overcome limited mass resolution, its fundamental assumption, that all satellite galaxies eventually merge with the central galaxy, is not consistent with the dynamics of subhalos at removal that we find, or with observed ICL fractions.

As an alternative, \citet{HenBerTho08}, using semi-analytic catalogs from the Millennium simulation, consider all satellite galaxies at $z=0$ whose subhalo has fallen below numerical resolution as tidally disrupting into ICL (implying that central-satellite mergers do not occur).
While this represents a highly simplified model, entirely contingent on a fixed numerical resolution mass, it does provide better agreement with the observed faint end of the galaxy luminosity function and allows for contribution to ICL consistent with observations.
\citet{KimBauCol09} use the same semi-analytic models and find better agreement with observed spatial clustering by assuming that satellite luminosity is instantly reduced at infall to roughly the ratio of satellite halo to host halo mass.
This model implies nearly total stellar mass stripping of low mass satellites immediately upon infall, in disagreement with our method which tracks mass loss over time across all satellite masses self-similarly.

Interestingly, \citet{MosSomMau09} use SHAM applied to a simulation of similar mass and force resolution to ours, and they find that they need to analytically track satellite galaxies after subhalo disruption to match galaxy clustering, in opposition to our findings.
This difference may arise because their simulation box size is $100\,h^{-1}$~Mpc, and as we discuss in the Appendix, simulation volumes smaller than $200\,h^{-1}$~Mpc tend to underpredict subhalo spatial clustering on all scales.

\section{Summary and Conclusion} \label{sec:summary}

We use a high-resolution dark matter-only $N$-body simulation of cosmological volume to track halos and their substructure, examining the fates of satellite galaxies.
Under the assumption that galaxies reside at the centers of subhalos and that stellar mass is related to infall mass, we assign stellar masses to subhalos through abundance matching, such that we recover the observed stellar mass function.
Our subhalo catalog intrinsically incorporates satellite-satellite mergers and satellite orbiting beyond their halo radius.
Thus, we examine different criteria for satellite galaxy removal (tidal disruption or merging with the central galaxy), using only information from the dark matter density field, and focusing primarily on the criterion based on the subhalo bound mass to infall mass ratio, $f_{\rm inf}=M_{\rm bound}/M_{\rm inf}$.
We highlight our main results as follows:

\begin{itemize}
\item Raising the threshold for removal, $f_{\rm inf}$, causes a reduction in the satellite HOD at all masses, such that more halos host only one galaxy.
Additionally, higher $f_{\rm inf}$ leads to a shallower HOD slope, implying that modelling satellite removal is more important in higher mass halos.
\item Disrupted/merged satellites reside primarily in the inner halo regions, and raising the removal threshold leads to a less concentrated satellite density profile.
A small fraction of satellites are disrupted out to the halo virial radius.
\item Most satellites fall below the removal threshold at a halo radius of $40-100\,h^{-1}$~kpc, the distance increasing with $f_{\rm inf}$.
The orbits of these satellites are only mildly radial and nearly half are directed outward from halo center, indicating that a substantial fraction of satellites are disrupted into the ICL as opposed to merging with the central galaxy.
\item Our subhalo catalog matched in number density to observed galaxy samples at $z=0.1$ reproduces well the projected correlation function at all scales and over a large range in subhalo masses using $f_{\rm inf}=0.01-0.03$.
Using reasonable ($0.2$ dex) luminosity-mass scatter does not change our results appreciably.
\item Our subhalo catalog also agrees well with observed satellite fractions and cluster satellite luminosity functions for similar values of $f_{\rm inf}$.
A scenario in which satellite galaxies never merge or disrupt is not feasible.
\item Methods for satellite removal based on physical separation from central galaxy or tidal radii predict too shallow of a dependence on subhalo mass.
Better agreement arises from methods that scale linearly with subhalo mass/size.
\item A simple analytical model for satellite removal timescale, based only on the satellite-halo mass ratio at infall and scaling with the Hubble time, reproduces well both our subhalo catalog up to high redshift, and the observed satellite fraction at $z=0.1$.
The best-fit $C_{dyn} \approx 0.25$ implies a substantially longer removal time than that derived from other dynamical friction models.
\end{itemize}

We emphasize that our results are based on examining subhalo merging/disruption in dark matter-only simulations, and the best-fit criteria to observations may change somewhat for simulations incorporating hydrodynamics and accurate galaxy stellar profiles.
For example, gas dynamics can affect orbits of the subhalos themselves, though this appears to be small \citep{JiaJinLin09}.
More importantly, a realistic subhalo stellar component is likely to be more compact and immune to stripping than the dark matter.
Recent works find some enhancement of satellite subhalo survivability in SPH simulations with star formation than in dark matter-only simulations, though the strength of this effect is unclear \citep{DolBorMur09,WeiColDav08}.

The largest source of uncertainty in our method is the assignment of stellar mass/luminosity to subhalos.
While abundance matching to subhalo infall mass does allow for ``passive'' satellite star formation as the $M_{\rm star}-M_{\rm inf}$ relation evolves, it may underpredict the star formation in satellite galaxies, particularly at higher redshift.
We have investigated how additional satellite star formation after infall influences our results by increasing all satellite infall masses by a fixed fraction.
Because our removal criterion is based on mass stripping, this is largely degenerate with increasing removal threshold.
Thus, to the extent that satellites experience significant star formation after infall (beyond what is implicitly in SHAM), our best-fit removal threshold may represent a slightly underestimate.
A more detailed prescription for star formation quenching can be related to environmental dependence of galaxy color, though we save incorporating properties beyond stellar mass or luminosity to future work.

Despite these uncertainties, our best fit mass stripping criterion for removal provides a consistent picture with galaxy stellar mass to dark mass ratios.
Abundance matching at $z=0.1$ gives $M_{\rm star}/M_{\rm sub,inf}\approx4\%$ for $M_{\rm inf}\sim10^{12}\,h^{-1}M_\odot$ (Fig.~\ref{fig:mstel}).
This implies that all of the stellar mass is only slightly more compact than the most bound dark matter of the same mass.
Thus, it is unlikely that galaxies within subhalos that fall below $f_{\rm inf}\sim0.01$ can remain intact.
If satellite galaxies have been stripped of most of their gas, then the gravitational dynamics of the stellar material will simply follow that of the dark matter, and since the dark mass fraction within a galaxy's radius is close to unity, dark mass stripping below $f_{\rm inf}\sim0.01$ must correspond to some amount of baryonic stripping.
While our binary model of merging/disruption is clearly an oversimplification, it does provides a good match to current observations.

Finally, our results imply strong constraints on simulations needed for robust subhalo populations, both in resolution and volume.
Taking $f_{\rm inf}=0.01$ as a conservative limit of our removal threshold and $\sim30$ particles as minimum resolution for tracking subhalos in dense environments, this implies that at least $\sim3000$ particles are needed at infall to robustly track satellites, which is consistent with our convergence tests on subhalo mass functions at $z=0.1$.
This resolution requirement is relaxed somewhat at high redshift ($z>1$), where average satellite times since infall are shorter so satellite have experienced less stripping.
Additionally, our tests on simulation volumes indicate that box sizes $<200\,h^{-1}$~Mpc have considerable difficulty reproducing representative halo mass functions and spatial clustering.
Taken together, these resolution and volume effects necessitate a wide dynamical range, using several billion particles, for simulations used to investigate galaxy evolution and galaxy large-scale structure.

\section*{Acknowledgments}

We thank J. Tinker for providing SDSS clustering data and 2dFGRS satellite fractions, A. Coil for DEEP2 clustering data, Z. Zheng for SDSS satellite fractions, and J. Cohn, C. Conroy, and J. Tinker for insightful comments on an early draft.
A.W. gratefully acknowledges the support of an NSF Graduate Research Fellowship and M.W. support from NASA and the DOE.
The simulations were analyzed at the National Energy Research Scientific 
Computing Center.

\appendix
\section{Impact of Simulation Size and Cosmology}

Given finite computational capacity, there is always a trade-off between simulation resolution and volume.
While our high-resolution $200\,h^{-1}$~Mpc simulation is able to track halo substructure with high fidelity, its limited volume poses problems for accurately recovering the high mass end of the halo mass function and large-scale spatial clustering.
This is not merely an issue of limited statistics and sample variance: by imposing mean density on a limited volume, we are not probing an entirely representative realization of the Universe, and the limited volume truncates power on scales larger than the box size.

Comparing the halo statistics of the $200\,h^{-1}$~Mpc and $720\,h^{-1}$~Mpc simulations (which have different resolution but the same cosmology and halo finder, as described in \S\ref{sec:simulations}), we find that the halo mass function at $z=0.1$ in the $200\,h^{-1}$~Mpc simulation is $5\%$ lower than that in the $720\,h^{-1}$~Mpc simulation from $3\times10^{11}\,h^{-1}M_\odot$ (the resolution limit of the larger simulation) to $3\times10^{14}\,h^{-1}M_\odot$, with no dependence on mass in this interval.
This implies that the deficit is driven not by resolution effects, but instead by truncated large-scale power and sample variance.
Above $3\times10^{14}\,h^{-1}M_\odot$, the $200\,h^{-1}$~Mpc simulation exhibits a significant deficit in halo density ($\sim20\%$), with its most massive halo being $8\times10^{14}\,h^{-1}M_\odot$, as opposed to $2\times10^{15}\,h^{-1}M_\odot$ in the larger simulation.
Additionally, we find an error of several percent in the halo correlation function measured at $10\,h^{-1}$~Mpc in the $200\,h^{-1}$~Mpc simulation as compared with the larger one, and it is significantly truncated on scales $\gtrsim 0.1\,L_{\rm box}$.

To test these finite-volume issues on subhalo clustering, we create a second subhalo catalog by mapping our high-resolution subhalo catalog onto the halo catalog of the larger, less resolved simulation.
Based on convergence tests from simulations of multiple sizes, the larger simulation accurately reproduces the spatial clustering on the scales of interest.
We randomly match two halos of the same mass from the two simulations, and we take subhalos from the halo in the high-resolution simulation and paste them onto the halo in the larger simulation.
We place the central subhalo at the potential minimum of the halo, and we place the satellite subhalos using a random overall orientation such that their positions with respect to all other subhalos in the host halo are retained.
Since halos of $M>8\times10^{14}\,h^{-1}M_\odot$ do not have a counterpart in the high-resolution simulation, we use fits to the satellite HOD extrapolated to higher halo mass to populate these halos in the larger simulation, assigning the satellite radial distributions to follow their halo NFW profiles with a random phase.
To examine the influence of cosmology, we also populate our subhalo catalog into our $500\,h^{-1}$~Mpc simulation which uses $n=1.0$ and $\sigma_8=0.9$.

\begin{figure}
\begin{center}
\resizebox{3.3in}{!}{\includegraphics{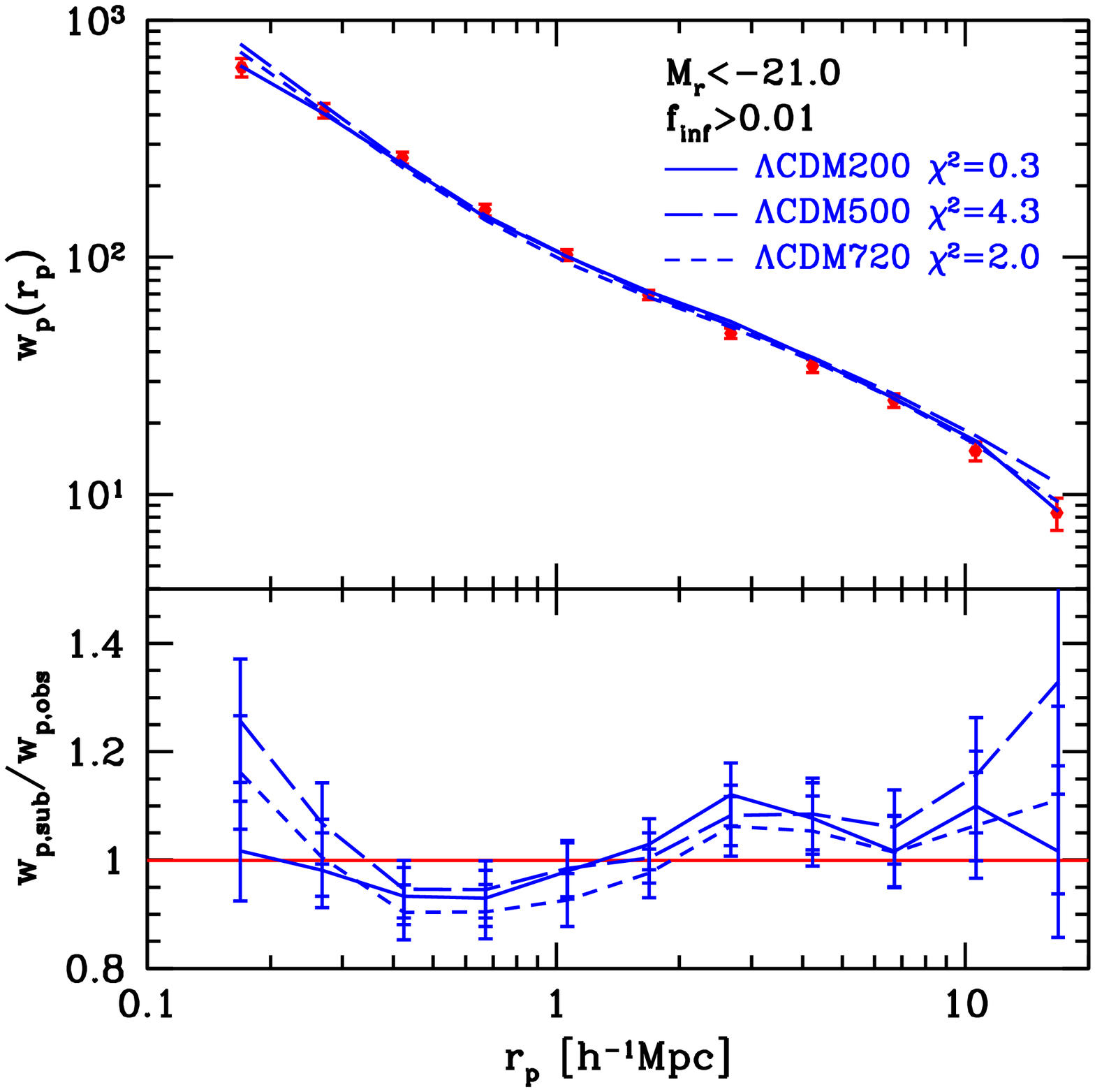}}
\end{center}
\vspace{-0.1in}
\caption{
Same as Fig.~\ref{fig:wp}, but using a fixed $f_{\rm inf}=0.01$ for subhalos directly from the $200\,h^{-1}$~Mpc simulation, and for subhalos populated into host halos in the $720\,h^{-1}$~Mpc simulation (same cosmology) and $500\,h^{-1}$~Mpc simulation (different cosmology).
$w_p(r_p)$ (top) and ratio of simulations to observed $w_p(r_p)$ (bottom).
Also shown is the reduced $\chi^2$ of the fit to observation for each simulation.
} \label{fig:wptest}
\end{figure}

Figure~\ref{fig:wptest} shows the effects of both simulation size and cosmology on the correlation function of subhalos corresponding to $M_r<-21.0$.
Subhalos in the larger simulation with the same cosmology exhibit somewhat enhanced $w_p(r_p)$ on the largest and smallest scales.
The change on moderate to small scales arises primarily because our re-population method ignores any environmental effects on the satellites, like halo alignment.
However, the large scale clustering amplitude is little affected.
We thus conclude that the $200\,h^{-1}$~Mpc simulation is not significantly affected by box size on these scales.

We emphasize, however, that the spatial clustering in smaller simulation box sizes is significantly affected on these scales.
We have performed the same comparisons with observations as in Fig.~\ref{fig:wptest} with a simulation of the same mass resolution and subhalo finder as the $200\,h^{-1}$~Mpc simulation, but was a smaller $125\,h^{-1}$~Mpc box\footnote{
The initial conditions for this simulation did had a some low modes at low $k$.}
We find a noticeable deficit in $w_p(r_p)$ at all subhalo masses both on large \textit{and} small scales as compared with observations and our $200\,h^{-1}$~Mpc box.
The former is readily expected, driven by mean density and a truncation of large-scale power.
The latter effect is more subtle, arising from the truncation of the halo mass function at high mass, which strongly affects the overall satellite population since these halos host many satellites.
Thus, we stress that in comparing the spatial clustering of subhalos in simulations to observed galaxies, both resolution and finite volume effects need be considered.

Figure~\ref{fig:wptest} also shows that changing cosmology to $n=1.0$ and $\sigma_8=0.9$ leads to a similar but stronger enhancement in $w_p(r_p)$ which is less consistent with observations.
Thus, if $\sigma_8$ is considerably larger than our fiducial value of $0.8$, our best-fit $f_{\rm inf}$ of $0.01$ might be an underestimate.

\bibliography{ms}

\label{lastpage}

\end{document}